\documentclass[aps,prl,reprint,superscriptaddress,nolongbibliography,floatfix]{revtex4-2}
\usepackage{amsmath,amssymb,bm,graphicx,booktabs}
\usepackage{xcolor}
\usepackage[T1]{fontenc}
\usepackage{tgtermes}
\usepackage[colorlinks,allcolors=blue]{hyperref}
\newcommand{\Rm}{r_{\mathrm{mass}}}

\newcommand{\dt}{\Delta\tau}
\newcommand{\tproj}{t_{\mathrm{proj}}}
\newcommand{\newsect}[1]{\noindent \textit{\textcolor{blue}{#1.---}}}
\usepackage[normalem]{ulem}
\begin{document}

\title{Single-Cone Chiral Ising Universality from Quantum Monte Carlo on the Sphere}
\author{Yuan Da Liao}
\author{Zi Yang Meng}
\affiliation{Department of Physics and HK Institute of Quantum Science \& Technology, The University of Hong Kong, Pokfulam Road, Hong Kong SAR, China}
\affiliation{State Key Laboratory of Optical Quantum Materials, The University of Hong Kong, Pokfulam Road, Hong Kong SAR, China}
\date{\today}

\begin{abstract}
How does a single Dirac cone acquire a mass through spontaneous Ising symmetry breaking? Quantum Monte Carlo (QMC) on the SLAC lattice established a benchmark for this transition, yet spherical spectra on 12 and 24 orbitals yielded markedly smaller inverse correlation-length exponents, $1/\nu=0.65$ and $0.79$, respectively. We present, to our knowledge, the first large-scale projector QMC study of single-cone chiral Ising Gross--Neveu--Yukawa criticality on the sphere. Our joint layer updates rebuild the propagator once per imaginary-time layer instead of once per field, reducing the computational burden by a factor of order $N_0$ and enabling simulations with 612 orbitals, 25.5 times the orbital count of the largest spherical exact-diagonalization study. From spherical correlations alone we obtain $1/\nu=1.23(7)$, $\eta_\phi=0.37(2)$, and $\eta_\psi=0.119(12)$. The mass-ratio scaling favors the SLAC estimates over the small-sphere spectral values, and our $1/\nu$ agrees with the mixed-correlator bootstrap. Our large-scale spherical QMC results establish an independent numerical benchmark and support a common universality class for the spherical and SLAC realizations.
\end{abstract}
\maketitle

\raggedbottom

\newsect{Introduction}
A single massless Dirac cone can emerge at a two-dimensional Chern-insulator phase transition, as in the Haldane model~\cite{Haldane1988}, or as a surface state of a three-dimensional topological insulator~\cite{Zhang2009Bi2Se3}. Angle-resolved photoemission observed these surface cones in Bi$_2$Se$_3$~\cite{Xia2009SingleCone}, Bi$_2$Te$_3$~\cite{Chen2009Bi2Te3}, TlBiSe$_2$~\cite{Sato2010TlBiSe2,Kuroda2010TlBiSe2} and TlBiTe$_2$~\cite{Chen2010TlBiTe2}. Spin-resolved measurements further revealed the surface state's helical spin texture~\cite {Hsieh2009Helical}. These observations raise the question of the fate of a single Dirac cone in the presence of interactions. When interactions spontaneously break an Ising symmetry, the resulting order parameter generates a fermion mass and opens a gap at the Dirac point. The minimal chiral Ising Gross--Neveu--Yukawa (GNY) theory -- the framework for interacting Dirac criticality~\cite{GrossNeveu1974,Herbut2006,Rosa2001FRG,Zerf2017FourLoop,Liu2020Designer,Wang2023EMUS} -- describes this process with one two-component complex Dirac field coupled to one real scalar~\cite{Tabatabaei2022,Erramilli2023}. Its critical exponents $\{\nu,\eta_\phi,\eta_\psi\}$ encode three distinct physical predictions: how rapidly a correlation length develops upon tuning the interaction, how the incipient mass order is correlated in space, and how the fermion propagates at criticality. Determining these exponents across different microscopic realizations provides a stringent test of universality for a single Dirac cone, whose regularization is constrained by fermion doubling and the parity anomaly~\cite{NielsenNinomiya1981,Redlich1984}.

On a square lattice with periodic boundary conditions, the SLAC derivative~\cite{Drell1976SLAC,rabinPerturbation1981,pelissettoLattice1988} realizes a single Dirac cone through long-range hopping. An on-site Hubbard interaction drives the transition to Ising mass order~\cite{Tabatabaei2022}. Tabatabaei et al.~\cite{Tabatabaei2022} extracted all three exponents by quantum Monte Carlo (QMC). Yu et al.~\cite{Yu2026SA} obtained compatible values through a different scaling analysis of the same model. Spherical geometry offers another route to quantum criticality~\cite{Zhu2023FuzzySphere,Hu2023SphereOPE,Zhou2024SphereSO5,ChenPhase2024,chenEmergent2024,Liao2026SO5WZW,Huffman2026SphereQCD}. For a single Dirac cone, spherical exact diagonalization (ED) identified low-lying excitations with the operator content expected at the GNY fixed point~\cite{Gao2026Sphere}. However, after tuning the interaction to resolve the conformal spectrum, the lowest even-scalar excitation gave $1/\nu\simeq0.65$ on 12 orbitals and $0.79$ on 24, well below the SLAC estimates near $\sim 1.2$. This discrepancy raises a question that small-sphere spectra alone cannot resolve: do correlations on larger spheres exhibit the same critical scaling as in the SLAC model?

To answer this question, we use projector QMC (PQMC) to study the spherical model with up to 612 orbitals, 25.5 times the largest orbital count in the spherical ED study. Our joint layer updates rebuild the propagator once per imaginary-time slice rather than once per field update, reducing the work per slice by a factor of order $N_0$, where $N_0$ counts the orbitals in each energy branch. From the mass ratio and the scalar and fermion correlations on the sphere, we independently determine the inverse correlation-length exponent $1/\nu$ and both anomalous dimensions $\eta_\phi$ and $\eta_\psi$. Our $1/\nu$ agrees with both SLAC estimates~\cite{Tabatabaei2022,Yu2026SA} and the mixed-correlator bootstrap~\cite{Erramilli2023}; together with the anomalous dimensions, this agreement supports a common universality class for the spherical and SLAC realizations. Our large-scale spherical PQMC results thus establish an independent numerical benchmark alongside the SLAC results for future studies of single-cone chiral Ising GNY criticality.
\par
\newsect{Model}
We start from the spherical Dirac Hamiltonian $H=H_0+H_{\rm int}$ of Ref.~\cite{Gao2026Sphere}, with
\begin{align}
 H_0&=\int d\Omega\,\Psi^\dagger(\Omega)
 \left[\frac{v}{R}(\bm\sigma\cdot\bm L+1)+h\sigma_z\right]
 \Psi(\Omega),\nonumber\\
 H_{\rm int}&=-\frac12\sum_{l\ge0}U_l\!
 \int d\Omega_a\,d\Omega_b\,m_{\Omega_a}m_{\Omega_b}\bigl[\nabla_{\Omega_a}^{(2l)}\delta(\Omega_a,\Omega_b)\bigr].
 \label{eq:ed-model}
\end{align}
At $\Omega=(\theta,\phi)$, the radial mass density is $m_\Omega=\Psi^\dagger(\Omega)(\hat{\bm n}_\Omega\cdot\bm\sigma)\Psi(\Omega)$, with outward unit normal $\hat{\bm n}_\Omega$ and solid-angle element $d\Omega=\sin\theta\,d\theta\,d\phi$. In Eq.~\eqref{eq:ed-model}, $\bm L$ is orbital angular momentum, and $\nabla_{\Omega_a}^{(2l)}$ differentiates the $d\Omega$-normalized spherical delta function to order $2l$ in $\Omega_a$. We set $v=1$ and $h=0$, retaining only $U_0=g/R^2$.
The delta function then sets $\Omega_a=\Omega_b$, giving $H_{\rm int}=-g/(2R^2)\int d\Omega\,m_\Omega^2$.

The retained $l=0$ interaction preserves rotations because $m_\Omega$ is a scalar under joint spin and spatial rotations. It conserves charge and is even under time reversal, parity, and charge conjugation~\cite{Gao2026Sphere}. We retain complete angular-momentum multiplets in both energy branches.

In the orbital basis, $\Psi(\Omega)=\sum_{skm}u_{skm}(\Omega)c_{skm}$, where $c_{skm}$ annihilates a fermion in the normalized eigenstate $u_{skm}$ of $h_0=(v/R)(\bm\sigma\cdot\bm L+1)$ with energy $svk/R$. Here $s=\pm$ labels the branch, $k=j+1/2=1,\ldots,n_{\rm cut}$ the shell, and $m=-j,\ldots,j$ its angular-momentum projection. With $N_0=n_{\rm cut}(n_{\rm cut}+1)$ orbitals per branch, the single-particle space has dimension $D=2N_0$. The 12- and 24-orbital ED systems thus have $N_0=6$ and 12, respectively~\cite{Gao2026Sphere}. We use half-filling and set the radius $R=\sqrt{N_0/2}$. The mass vertex for a real orthonormal spherical harmonic $Y_{\ell\mu}$ is
\begin{equation}
 (A_{\ell\mu})_{skm,s'k'm'} =
 \int d\Omega\,Y_{\ell\mu}(\Omega) u_{skm}^\dagger(\Omega)
 (\hat{\bm n}_\Omega\cdot\bm\sigma)u_{s'k'm'}(\Omega).
 \label{eq:vertex-main}
\end{equation}
Equation~\eqref{eq:sm-gaunt-vertex} in the Supplemental Material (SM)~\cite{SupplementalMaterial} evaluates these matrix elements. The orbital Hamiltonian then reads
\begin{align}
 H_0 &= \sum_{skm}\frac{svk}{R}c_{skm}^\dagger c_{skm},
 \label{eq:kinetic-main}\\
 H_{\rm int} &= -\frac{g}{2R^2}\sum_{\ell=0}^{2n_{\rm cut}-1}
 \sum_{\mu=-\ell}^{\ell} Q_{\ell\mu}^2,
 \label{eq:interaction-main}
\end{align}
where the mass multipoles are
\[
\begin{aligned}
 Q_{\ell\mu} &= \int d\Omega\,Y_{\ell\mu}(\Omega)m_\Omega \\
 &=\sum_{skm,s'k'm'}c_{skm}^\dagger
 (A_{\ell\mu})_{skm,s'k'm'}c_{s'k'm'}.
\end{aligned}
\]
The monopole ($\ell=0$) measures the uniform mass, and higher multipoles resolve its angular variation. Angular-momentum addition gives $\ell_{\max}=2j_{\max}=2n_{\rm cut}-1$. Our largest computationally feasible scan reaches $n_{\rm cut}=17$, corresponding to $N_0=306$ and $\ell_{\max}=33$.

\newsect{Method}
The fermion weights in our PQMC simulations exhibit a sign problem~\cite{Troyer2005,panSign2024}, which we handle by sampling their moduli and retaining the full complex phase in measurements. With a centered imaginary-density Hubbard--Stratonovich (HS) decoupling, the phase signal near the transition remains readily resolvable up to $N_0=306$ (612 orbitals), 25.5 times the largest orbital count in the spherical ED study~\cite{Gao2026Sphere} [see Fig.~\ref{fig:sm-fig1-phase} in the SM~\cite{SupplementalMaterial}].

We construct the trial state $|\Psi_T\rangle$ by filling all $N_0$ negative-energy orbitals of $H_0$, forming the free Dirac sea at half filling. Projecting this state by $\Theta/2$ on each side of the observable gives
\begin{equation}
 \langle O\rangle_\Theta=
 \frac{\langle\Psi_T|e^{-\Theta H/2}Oe^{-\Theta H/2}|\Psi_T\rangle}
 {\langle\Psi_T|e^{-\Theta H}|\Psi_T\rangle}.
 \label{eq:projector}
\end{equation}
There are $L_\tau$ time slices with $\dt=\Theta/L_\tau$. The imaginary-density scan uses $\tproj=\Theta/R=0.6$ and a time step $\dt\le0.1$. We assume $z=1$ in the finite-size analysis.

The centered imaginary-density decoupling follows from the exact cutoff identity $m_\Omega^2=\kappa^2-(\rho_\Omega-\kappa)^2$, where $\rho_\Omega=\Psi^\dagger\Psi$ and $\kappa=N_0/(4\pi)$. In the half-filled sector, the centered density vertex is $\bar T_q=T_q-\operatorname{tr}(T_q)I_D/D$, where $q=(\ell,\mu)$ and
\begin{equation}
 (T_{\ell\mu})_{skm,s'k'm'}=\int d\Omega\,Y_{\ell\mu}(\Omega)
 u_{skm}^\dagger(\Omega)u_{s'k'm'}(\Omega).
\end{equation}
The evaluated matrix elements are given in Eq.~\eqref{eq:sm-gaunt-density} in the SM~\cite{SupplementalMaterial}. For the $t$-th imaginary-time slice, the single-particle propagator is
\begin{equation}
 \begin{aligned}
 B_t&=e^{-\dt K/2}e^{iW_t}e^{-\dt K/2},\\
 W_t&=\sqrt{\dt g/R^2}\sum_{q\ne00}x_{qt}\bar T_q .
 \end{aligned}
 \label{eq:density-layer-main}
\end{equation}
Here $K$ is the one-particle kinetic matrix, diagonal in $(s,k,m)$ with entries $svk/R$, and $x_{qt}$ is the real Gaussian auxiliary field for density multipole $q$ on slice $t$.
The uniform centered-density vertex vanishes at half filling, and $e^{iW_t}$ is unitary. SM~\cite{SupplementalMaterial} gives the decoupling derivation and the implementation details.

At $N_0=6$ and $g=15.1$, the PQMC energy and scalar structure factor $S_0$ approach the exact-diagonalization values at the same projection length as $\Delta\tau$ decreases [Fig.~\ref{fig:ed-dtau} in the SM~\cite{SupplementalMaterial}]. 

\newsect{Joint layer updates}
On a finite sphere, the lowest positive and highest negative energies are $\pm v/R$. Every negative-energy shell is filled, so degeneracy within a shell leaves the occupied subspace unique. Let $Q_T$ be the $2N_0\times N_0$ matrix whose columns are the occupied negative-energy orbitals of the free Dirac sea. The Gaussian field measure is $P[X]=\prod_{qt}p(x_{qt})$, with $p(x)=e^{-x^2/2}/\sqrt{2\pi}$ for each active field. The weight $w[X]=P[X]\mathcal D[X]$ includes the Slater overlap
\begin{equation}
 \mathcal D[X]=\det[Q_T^\dagger B_{L_\tau}\cdots B_1Q_T].\label{eq:weight-main}
\end{equation}
The factor $e^{\Theta C_0}$ cancels in weight ratios and normalized measurements. At slice $t$, let $L_t^\dagger=Q_T^\dagger B_{L_\tau}\cdots B_{t+1}$ and $R_t=B_{t-1}\cdots B_1Q_T$. Replacing $B_t$ by $B_t'$ gives the overlap ratio $r_D=\det(L_t^\dagger B_t'R_t)/\det(L_t^\dagger B_tR_t)$.

We update all active fields on a single imaginary-time layer jointly, using a Metropolis-adjusted Langevin proposal. Related continuous-field updates have been used in a projected SO(5) model~\cite{Liao2026SO5WZW} and in twisted bilayer graphene~\cite{Huang2025AngleGN,shanTrion2026}. For each active field on slice $t$, we propose
\begin{equation}
 x'_{qt}=x_{qt}+\frac{b^2}{2}[F_q(X)-x_{qt}]+b\,\xi_q,
 \qquad \xi_q\sim\mathcal N(0,1).
 \label{eq:density-proposal-main}
\end{equation}
Here $F_q$ is a sparse approximation to the determinant drift (see Ref.~\cite{Shi2015ForceBias} for an auxiliary-field force-bias precedent), and $-x_{qt}$ is the Gaussian-prior drift. We tune $b$ during warmup following the strategy of the projected SO(5) study~\cite{Liao2026SO5WZW} and hold it fixed thereafter. We evaluate the reverse proposal at the candidate and accept the whole layer with
\begin{equation}
 a(X\!\to\!X')=\min\!\left\{1,
 |r_D|\frac{P[X']}{P[X]}
 \frac{\mathcal Q(X|X')}{\mathcal Q(X'|X)}\right\}.\label{eq:mh-main}
\end{equation}
The joint proposal density $\mathcal Q$ is the product of the Gaussian densities in Eq.~\eqref{eq:density-proposal-main}. The reverse drift is evaluated at $X'$. The Metropolis--Hastings correction targets $|w|$ for the finite-time-step transfer operator~\cite{Hastings1970}. SM~\cite{SupplementalMaterial} explains the force $F_q$ computation and warmup rule.

Rebuilding the dense propagator and Slater overlap after every channel change costs $O(N_qD^3)$ per layer. A joint layer update constructs one candidate propagator and evaluates one overlap ratio. The selection rules restrict assembly to $O(D^2n_{\rm cut})$ operations. With $D=2N_0$ and $N_q=O(N_0)$, the leading cost per layer falls from $O(N_0^4)$ to $O(N_0^3)$. This reduction in the cost per layer helped us reach $2N_0=612$ orbitals, 25.5 times the largest orbital count in the spherical ED study~\cite{Gao2026Sphere}.

\begin{figure}[!t]
\centering
\includegraphics[width=\columnwidth]{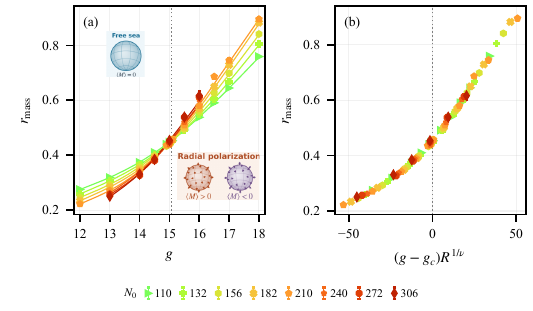}
\caption{\textbf{Critical scaling from PQMC.} (a) Mass ratio for $110\le N_0\le306$ and $12\le g\le18$. Colored curves are separate smoothing splines for each size. Insets show the free sea and the two radial-polarization orientations. (b) Mass-ratio collapse for $N_0=110,132,156,182,210,240,272,306$ and $12\le g\le18$. The inverse-variance fit gives $g_c=15.07(2)$ and $1/\nu=1.23(7)$. The dotted vertical lines mark $g_c$ in (a) and $x=0$ in (b).}
\label{fig:results}
\end{figure}

\newsect{Crossings and the correlation-length exponent}
To extract the inverse correlation-length exponent directly from correlations, we first locate the transition with $\Rm=1-S_1/S_0$, where $S_\ell=(2\ell+1)^{-1}\sum_\mu\langle Q_{\ell\mu}^2\rangle_\Theta$. At $\tproj=0.6$, the size ordering reverses across $12\le g\le18$ [Fig.~\ref{fig:results}(a)]. The mass-ratio crossings drift toward lower couplings as the size increases. An inverse-variance fit of $\Rm$ over $110\le N_0\le306$ and $12\le g\le18$ determines $g_c$ and $1/\nu$ jointly. The collapse of $\Rm$ as a function of $x=(g-g_c)R^{1/\nu}$ yields $g_c=15.07(2)$ and $1/\nu=1.23(7)$ [Fig.~\ref{fig:results}(b)]. The exponent uncertainties include the observed size- and coupling-window dependence; the uncertainty on $g_c$ is statistical (see the SM~\cite{SupplementalMaterial} for both estimates). Our estimate agrees within uncertainties with both SLAC QMC results, $1.19(3)$ and $1.18(3)$~\cite{Tabatabaei2022,Yu2026SA}. Our result also agrees within uncertainties with the mixed-correlator bootstrap range, $1/\nu=3-\Delta_\epsilon=1.26(4)$~\cite{Erramilli2023}, and lies substantially above the spherical ED estimates of $0.65$ and $0.79$~\cite{Gao2026Sphere} (Table~\ref{tab:compare}).

\begin{figure}[t]
\centering
\includegraphics[width=\columnwidth]{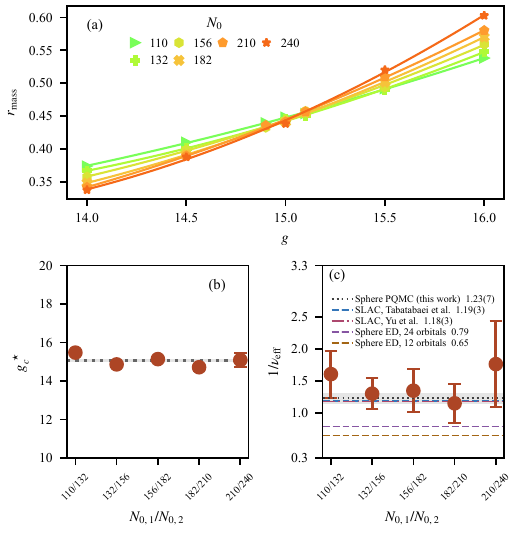}
\caption{\textbf{Local crossing and slope tests.} (a) Mass ratios for the six sizes from $N_0=110$ to $240$ and their equal-weight quadratic fits over $14\le g\le16$. (b) Crossing couplings $g_c^\star$ for adjacent sizes from the quadratic fits. (c) Finite-size estimates $1/\nu_{\rm eff}$ from the slopes of the same fits at each crossing [Eq.~\eqref{eq:crossing-slope}]. Colored reference lines in (c) mark the SLAC QMC central values $1.19$ (blue dashed) and $1.18$ (rose dash-dotted)~\cite{Tabatabaei2022,Yu2026SA} and the spherical ED estimates $0.79$ (purple, 24 orbitals) and $0.65$ (ochre, 12 orbitals)~\cite{Gao2026Sphere}. Dotted lines and shaded bands in (b,c) show the global estimates from Fig.~\ref{fig:results} and their quoted uncertainties, $g_c=15.07(2)$ and $1/\nu=1.23(7)$, respectively. Pointwise error bars give one standard deviation from whole-chain bootstrap replicates with a unique crossing in the fit window.}
\label{fig:local-crossings}
\end{figure}

Figure~\ref{fig:local-crossings} tests the global collapse using separate quadratic fits to the mass-ratio curves over $14\le g\le16$. We use adjacent size pairs between $N_0=110$ and $240$, where the coupling grid near the transition is fine enough to resolve both the crossing and the local slopes. Following the crossing-point analysis of Refs.~\cite{ChenPhase2024,qinDuality2017}, we locate the intersection of the two fitted mass-ratio curves for each pair of adjacent radii $R_1<R_2$. Its coupling $g_c^\star$ estimates $g_c$ [Fig.~\ref{fig:local-crossings}(b)]. The mass-ratio slope grows as $R^{1/\nu}$ to leading order, so the ratio of the slopes at the crossing gives [Fig.~\ref{fig:local-crossings}(c)]
\begin{equation}
 \frac{1}{\nu_{\rm eff}}=\frac{\ln(s_2/s_1)}{\ln(R_2/R_1)},\qquad
 s_i=\left.\partial_g\Rm(g,R_i)\right|_{g=g_c^\star}.
 \label{eq:crossing-slope}
\end{equation}
The finite-size corrections are derived in the SM~\cite{SupplementalMaterial}. The crossings in Fig.~\ref{fig:local-crossings}(b) scatter on both sides of the global estimate $g_c=15.07(2)$. The $156/182$ and $210/240$ pairs give $g_c^\star=15.13(13)$ and $15.09(34)$, respectively, in agreement with the collapse, although the pairwise crossings show no monotonic drift. Four of the five slope quotients in Fig.~\ref{fig:local-crossings}(c) agree within one bootstrap standard error with both SLAC central values, $1.19$ and $1.18$~\cite{Tabatabaei2022,Yu2026SA}, and with our global estimate, $1.23$. The $110/132$ pair lies higher, but its bootstrap error bar overlaps the shaded interval $1.16\le 1/\nu\le 1.30$ from the global collapse. Both spherical ED estimates, $0.79$ and $0.65$~\cite{Gao2026Sphere}, lie below all five error bars. Thus the local sharpening of the mass-ratio curves favors the larger inverse correlation-length exponent found in the QMC studies.

\begin{table}[!t]
\caption{\textbf{Single-cone critical exponents.} The last column reports $\chi^2$ per degree of freedom from our fits to the same spherical PQMC mass-ratio data ($\tproj=0.6$, $110\le N_0\le306$, $12\le g\le18$). We fit $1/\nu$ freely in the first row and fix it to the published central value in each SLAC and ED row. In every fit we optimize $g_c$ and the common scaling spline using the same inverse-variance weights [see Fig.~\ref{fig:sm-literature-collapse} in the SM~\cite{SupplementalMaterial}]. PQMC exponent errors include size- and coupling-window variation. For the $N=2$ bootstrap comparison, we quote the $\Delta_{\sigma'}>2.5$ results from Table~1 of Ref.~\cite{Erramilli2023}. The full set of spectral assumptions is given in Sec.~2.3 of that work. In the ED rows, $\eta_\phi=0.32$ is an input from the velocity calibration at $\Delta_\sigma=0.66$.}
\label{tab:compare}
\begin{ruledtabular}
\begin{tabular}{@{}lccc|c@{}}
Source & $1/\nu$ & $\eta_\phi$ & $\eta_\psi$ & $\chi^2/\mathrm{dof}$\\
\hline
Sphere PQMC (this work) & 1.23(7) & 0.37(2) & 0.119(12) & 2.88\\
SLAC QMC~\cite{Tabatabaei2022} & 1.19(3) & 0.31(1) & 0.136(5) & 2.98\\
SLAC QMC~\cite{Yu2026SA} & 1.18(3) & 0.36(3) & 0.134(3) & 3.06\\
Sphere ED, 24 orbitals~\cite{Gao2026Sphere} & 0.79 & 0.32 (input) & 0.14 & 24.04\\
Sphere ED, 12 orbitals~\cite{Gao2026Sphere} & 0.65 & 0.32 (input) & 0.10 & 40.67\\
Mixed bootstrap~\cite{Erramilli2023} & 1.26(4) & 0.313(25) & 0.134(5) & ---\\
\end{tabular}
\end{ruledtabular}
\end{table}

\begin{figure}[t]
\centering
\includegraphics[width=\columnwidth]{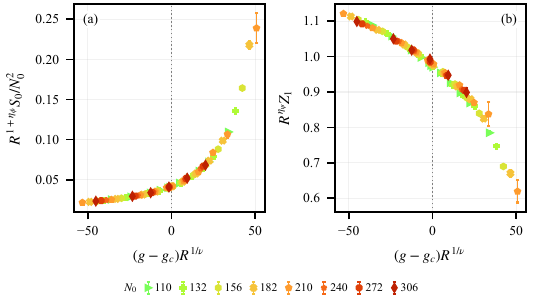}
\caption{\textbf{Anomalous dimensions.} The scalar structure factor $S_0$ in (a) and the lowest-shell fermion contrast $Z_1$ in (b) are rescaled using inverse-variance fits over $110\le N_0\le306$ and $12\le g\le18$, yielding $\eta_\phi=0.37(2)$ and $\eta_\psi=0.119(12)$, respectively. Both panels use the scaling coordinate $x=(g-g_c)R^{1/\nu}$ from the mass-ratio fit in Fig.~\ref{fig:results}(b). Dotted lines mark $x=0$.}
\label{fig:anomalous}
\end{figure}

\newsect{Scalar and fermion dimensions}
The mass structure factor probes the scalar field, while the shell contrast $Z_k=(2k)^{-1}\sum_m\langle n_{-km}-n_{+km}\rangle_\Theta$ probes the fermion. The free sea has $Z_k=1$. At fixed density, $S_0/N_0^2$ is proportional to the mean square of the spatially averaged mass. The lowest shell has momentum $1/R$, and normalized orbital projection of the one-body correlator gives $Z_1\propto R^{2-2\Delta_\psi}$. Thus
\begin{equation}
 \frac{S_0}{N_0^2}=R^{-(1+\eta_\phi)}F_\phi(x),\qquad
 Z_1=R^{-\eta_\psi}F_Z(x).\label{eq:anomalous-scaling}
\end{equation}
With the horizontal scaling coordinate fixed by the mass-ratio fit, inverse-variance fits give $\eta_\phi=0.37(2)$ and $\eta_\psi=0.119(12)$ [Fig.~\ref{fig:anomalous}].

\begin{figure}[t]
\centering
\includegraphics[width=\columnwidth]{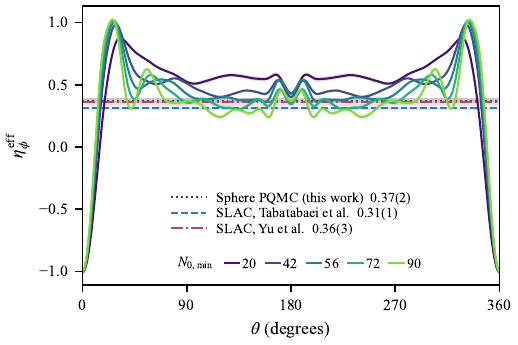}
\caption{\textbf{Real-space test of the scalar anomalous dimension at $g_c$.} The radius fits use $N_0=20,42,56,72,90,110,132,156,182,210$. At each angle we fit $\ln C_R(\theta)$ versus $\ln R$ with slope $-(1+\eta_\phi^{\rm eff})$. Each curve includes all sizes from the indicated lower cutoff to $N_0=210$, giving ten to six radii per fit. Colored bands show pointwise jackknife uncertainties. Angles beyond $180^\circ$ mirror the same physical separations. Horizontal lines and their shaded bands give the central values and quoted uncertainties in Table~\ref{tab:compare}: spherical PQMC, $0.37(2)$ (gray dotted), and SLAC QMC, $0.31(1)$ (blue dashed)~\cite{Tabatabaei2022} and $0.36(3)$ (rose dash-dotted)~\cite{Yu2026SA}.}
\label{fig:realspace-even}
\end{figure}

The spatial decay of mass correlations gives a second route to $\eta_\phi$. For the radius fits we choose $g=15.1$, a simulated coupling close to the fitted critical value $g_c=15.07(2)$. We reconstruct the physical equal-time correlator from the measured mass multipoles,
$C_R(\theta)=(4\pi R^4)^{-1}\sum_\ell(2\ell+1)S_\ell P_\ell(\cos\theta)$.
For a scalar primary of dimension $\Delta_\phi=(1+\eta_\phi)/2$, the equal-time conformal correlator on a sphere is proportional to $[2R\sin(\theta/2)]^{-2\Delta_\phi}$. We fit the radius dependence at fixed angle, allowing the amplitude to vary with angle. The predicted power law is $C_R(\theta)\propto R^{-(1+\eta_\phi)}$~\cite{Huffman2026SphereQCD}. A radius fit at each angle therefore tests the scalar exponent independently of the coupling-scan scaling function and its fitted $1/\nu$. Figure~\ref{fig:realspace-even} shows how the angular profile changes as smaller spheres are excluded. Averaging over $90^\circ\leq\theta\leq270^\circ$ gives $0.543(15)$ for all ten radii and $0.356(16)$ for $N_{0,\min}=72$, consistent in the latter window with the $S_0$ estimate. Increasing $N_{0,\min}$ to 90 lowers the average to $0.311(22)$, illustrating the sensitivity to the radius window.

\newsect{Comparison across regularizations}
In the projected SO(5) model, simulations on the torus and sphere regularizations found the consistent phase structure and exponents~\cite{Liao2026SO5WZW}. We perform a similar analysis here. The SLAC lattice realizes a single cone through long-range hopping, while the sphere retains complete angular-momentum shells. If both realizations flow to the same fixed point, changing the ultraviolet construction should preserve the scaling of the mass fluctuations and of the fermion field. We test both predictions using exponents fitted entirely to the spherical data. The mass structure factor gives $\eta_\phi=0.37(2)$, and the lowest-shell fermion contrast gives $\eta_\psi=0.119(12)$. The scalar estimate overlaps the result of Yu et al.~\cite{Yu2026SA}; our fermion estimate has a lower central value, but its uncertainty interval overlaps those of both SLAC estimates (Table~\ref{tab:compare})~\cite{Tabatabaei2022,Yu2026SA}. On the largest spheres, the fixed-angle radius fits yield an $\eta_\phi$ consistent with both our collapse estimate and the SLAC result of Yu et al.~\cite{Yu2026SA}, independently of the fitted $1/\nu$ [Fig.~\ref{fig:realspace-even}].

The spherical mass ratios distinguish our PQMC and the SLAC estimates of $1/\nu$ from the small-size ED estimate, even when $g_c$ and the scaling function are reoptimized for each exponent. The unconstrained fit gives $1/\nu=1.23(7)$ with a reduced $\chi^2$ of 2.88. Fixing the exponent to either SLAC value~\cite{Tabatabaei2022,Yu2026SA} raises it only to 2.98 or 3.06, whereas the 24- and 12-orbital ED values~\cite{Gao2026Sphere} give 24.04 and 40.67, respectively [Table~\ref{tab:compare} here and Fig.~\ref{fig:sm-literature-collapse} in the SM~\cite{SupplementalMaterial}]. Our estimate also agrees with the mixed-correlator bootstrap result of $1.26(4)$ for two Majorana components coupled to a single Ising scalar~\cite{Erramilli2023}.

On the sphere itself, ED inferred $1/\nu=3-\Delta_\epsilon$ from the lowest even-scalar excitation. Its estimate rose from 0.65 at 12 orbitals to 0.79 at 24 orbitals~\cite{Gao2026Sphere}, approaching the earlier fermion-only bootstrap estimate of about $0.86$~\cite{Iliesiu2018,Gao2026Sphere}. This upward drift points to finite-size effects as the most likely source of the low spectral estimates. The comparison also involves a change of Hamiltonian and observable: ED tuned derivative interactions to resolve the low-energy spectrum, whereas our PQMC calculation extracts the exponent from mass correlations with only the $l=0$ interaction, reaching 612 orbitals, 25.5 times the largest ED orbital count.

\newsect{Conclusion}
Large-scale spherical PQMC establishes a comprehensive numerical analysis for the minimal chiral Ising GNY transition. With 612 orbitals, 25.5 times the largest spherical ED orbital count, we can determine $1/\nu$ from the growth of correlations with radius. Joint layer updates make this size range accessible by treating all density channels in a single candidate propagator. This organization is applicable to other projected Hamiltonians with many coupled density modes, including Landau-level and moir\'{e}-band models~\cite{Liao2026SO5WZW,Huang2025AngleGN,shanMomentum2026,shanTrion2026}. The result, $1/\nu =1.23(7)$, agrees within uncertainties with both SLAC determinations and the mixed-correlator bootstrap and lies well above the two small-sphere ED estimates. The distinction survives a direct test against the mass ratios even when the critical coupling is refitted for each proposed exponent.

The two critical fields give complementary evidence. The scalar exponent $\eta_\phi=0.37(2)$ agrees with the SLAC estimate of Yu et al.~\cite{Yu2026SA}. The real-space estimate drifts with the radius window. The fermion exponent $\eta_\psi=0.119(12)$ has a slightly lower central value, with an uncertainty interval overlapping both SLAC estimates (Table~\ref{tab:compare}). The spherical and SLAC calculations~\cite{Tabatabaei2022,Yu2026SA} thus support a common infrared description of spontaneous Dirac mass generation. Just as the Ising exponents provide a reference for criticality driven by a single scalar field, the single-cone chiral Ising GNY transition provides a reference for criticality in which a mass order parameter and a Dirac fermion fluctuate together. Our large-scale spherical QMC results establish an independent benchmark for this minimal fermionic critical point against which theories of more complex Dirac transitions can be tested.

\begin{samepage}
\begin{acknowledgments}
\newsect{Acknowledgments} Y.D.L. acknowledges funding from the National Natural Science Foundation of China (Grant No.~12404282). Y.D.L. and Z.Y.M. acknowledge support from the Guangdong Natural Science Foundation General Program (Grant No.~2025A1515010337), the Hong Kong Research Grants Council (Project Nos. C7037-22GF,
17301924, 17301725, 17302026 and AoE/P-604/25-R), the joint scheme of the French National Research Agency and the Hong Kong Research Grants Council (Project No.~A\_HKU703/22), and the State Key Laboratory of Optical Quantum Materials at the University of Hong Kong. The authors gratefully acknowledge the computing time made available to them on the high-performance computer Barnard at the NHR Center of TU Dresden. This center is jointly supported by the Federal Ministry of Education and Research and the state governments participating in the National High-Performance Computing (NHR) joint funding program (\url{http://www.nhr-verein.de/en/our-partners}).
\end{acknowledgments}
\end{samepage}

\bibliography{references}

\begin{thebibliography}{48}%
\makeatletter
\providecommand \@ifxundefined [1]{%
 \@ifx{#1\undefined}
}%
\providecommand \@ifnum [1]{%
 \ifnum #1\expandafter \@firstoftwo
 \else \expandafter \@secondoftwo
 \fi
}%
\providecommand \@ifx [1]{%
 \ifx #1\expandafter \@firstoftwo
 \else \expandafter \@secondoftwo
 \fi
}%
\providecommand \natexlab [1]{#1}%
\providecommand \enquote  [1]{``#1''}%
\providecommand \bibnamefont  [1]{#1}%
\providecommand \bibfnamefont [1]{#1}%
\providecommand \citenamefont [1]{#1}%
\providecommand \href@noop [0]{\@secondoftwo}%
\providecommand \href [0]{\begingroup \@sanitize@url \@href}%
\providecommand \@href[1]{\@@startlink{#1}\@@href}%
\providecommand \@@href[1]{\endgroup#1\@@endlink}%
\providecommand \@sanitize@url [0]{\catcode `\\12\catcode `\$12\catcode
  `\&12\catcode `\#12\catcode `\^12\catcode `\_12\catcode `\%12\relax}%
\providecommand \@@startlink[1]{}%
\providecommand \@@endlink[0]{}%
\providecommand \url  [0]{\begingroup\@sanitize@url \@url }%
\providecommand \@url [1]{\endgroup\@href {#1}{\urlprefix }}%
\providecommand \urlprefix  [0]{URL }%
\providecommand \Eprint [0]{\href }%
\providecommand \doibase [0]{https://doi.org/}%
\providecommand \selectlanguage [0]{\@gobble}%
\providecommand \bibinfo  [0]{\@secondoftwo}%
\providecommand \bibfield  [0]{\@secondoftwo}%
\providecommand \translation [1]{[#1]}%
\providecommand \BibitemOpen [0]{}%
\providecommand \bibitemStop [0]{}%
\providecommand \bibitemNoStop [0]{.\EOS\space}%
\providecommand \EOS [0]{\spacefactor3000\relax}%
\providecommand \BibitemShut  [1]{\csname bibitem#1\endcsname}%
\let\auto@bib@innerbib\@empty
\bibitem [{\citenamefont {Haldane}(1988)}]{Haldane1988}%
  \BibitemOpen
  \bibfield  {author} {\bibinfo {author} {\bibfnamefont {F.~D.~M.}\
  \bibnamefont {Haldane}},\ }\href
  {https://doi.org/10.1103/PhysRevLett.61.2015} {\bibfield  {journal} {\bibinfo
   {journal} {Phys. Rev. Lett.}\ }\textbf {\bibinfo {volume} {61}},\ \bibinfo
  {pages} {2015} (\bibinfo {year} {1988})}\BibitemShut {NoStop}%
\bibitem [{\citenamefont {Zhang}\ \emph {et~al.}(2009)\citenamefont {Zhang},
  \citenamefont {Liu}, \citenamefont {Qi}, \citenamefont {Dai}, \citenamefont
  {Fang},\ and\ \citenamefont {Zhang}}]{Zhang2009Bi2Se3}%
  \BibitemOpen
  \bibfield  {author} {\bibinfo {author} {\bibfnamefont {H.}~\bibnamefont
  {Zhang}}, \bibinfo {author} {\bibfnamefont {C.-X.}\ \bibnamefont {Liu}},
  \bibinfo {author} {\bibfnamefont {X.-L.}\ \bibnamefont {Qi}}, \bibinfo
  {author} {\bibfnamefont {X.}~\bibnamefont {Dai}}, \bibinfo {author}
  {\bibfnamefont {Z.}~\bibnamefont {Fang}},\ and\ \bibinfo {author}
  {\bibfnamefont {S.-C.}\ \bibnamefont {Zhang}},\ }\href
  {https://doi.org/10.1038/nphys1270} {\bibfield  {journal} {\bibinfo
  {journal} {Nat. Phys.}\ }\textbf {\bibinfo {volume} {5}},\ \bibinfo {pages}
  {438} (\bibinfo {year} {2009})}\BibitemShut {NoStop}%
\bibitem [{\citenamefont {Xia}\ \emph {et~al.}(2009)\citenamefont {Xia},
  \citenamefont {Qian}, \citenamefont {Hsieh} \emph
  {et~al.}}]{Xia2009SingleCone}%
  \BibitemOpen
  \bibfield  {author} {\bibinfo {author} {\bibfnamefont {Y.}~\bibnamefont
  {Xia}}, \bibinfo {author} {\bibfnamefont {D.}~\bibnamefont {Qian}}, \bibinfo
  {author} {\bibfnamefont {D.}~\bibnamefont {Hsieh}}, \emph {et~al.},\ }\href
  {https://doi.org/10.1038/nphys1274} {\bibfield  {journal} {\bibinfo
  {journal} {Nat. Phys.}\ }\textbf {\bibinfo {volume} {5}},\ \bibinfo {pages}
  {398} (\bibinfo {year} {2009})}\BibitemShut {NoStop}%
\bibitem [{\citenamefont {Chen}\ \emph {et~al.}(2009)\citenamefont {Chen},
  \citenamefont {Analytis}, \citenamefont {Chu}, \citenamefont {Liu},
  \citenamefont {Mo}, \citenamefont {Qi}, \citenamefont {Zhang}, \citenamefont
  {Lu}, \citenamefont {Dai}, \citenamefont {Fang}, \citenamefont {Zhang},
  \citenamefont {Fisher}, \citenamefont {Hussain},\ and\ \citenamefont
  {Shen}}]{Chen2009Bi2Te3}%
  \BibitemOpen
  \bibfield  {author} {\bibinfo {author} {\bibfnamefont {Y.~L.}\ \bibnamefont
  {Chen}}, \bibinfo {author} {\bibfnamefont {J.~G.}\ \bibnamefont {Analytis}},
  \bibinfo {author} {\bibfnamefont {J.-H.}\ \bibnamefont {Chu}}, \bibinfo
  {author} {\bibfnamefont {Z.~K.}\ \bibnamefont {Liu}}, \bibinfo {author}
  {\bibfnamefont {S.~K.}\ \bibnamefont {Mo}}, \bibinfo {author} {\bibfnamefont
  {X.~L.}\ \bibnamefont {Qi}}, \bibinfo {author} {\bibfnamefont {H.~J.}\
  \bibnamefont {Zhang}}, \bibinfo {author} {\bibfnamefont {D.~H.}\ \bibnamefont
  {Lu}}, \bibinfo {author} {\bibfnamefont {X.}~\bibnamefont {Dai}}, \bibinfo
  {author} {\bibfnamefont {Z.}~\bibnamefont {Fang}}, \bibinfo {author}
  {\bibfnamefont {S.~C.}\ \bibnamefont {Zhang}}, \bibinfo {author}
  {\bibfnamefont {I.~R.}\ \bibnamefont {Fisher}}, \bibinfo {author}
  {\bibfnamefont {Z.}~\bibnamefont {Hussain}},\ and\ \bibinfo {author}
  {\bibfnamefont {Z.-X.}\ \bibnamefont {Shen}},\ }\href
  {https://doi.org/10.1126/science.1173034} {\bibfield  {journal} {\bibinfo
  {journal} {Science}\ }\textbf {\bibinfo {volume} {325}},\ \bibinfo {pages}
  {178} (\bibinfo {year} {2009})}\BibitemShut {NoStop}%
\bibitem [{\citenamefont {Sato}\ \emph {et~al.}(2010)\citenamefont {Sato},
  \citenamefont {Segawa}, \citenamefont {Guo}, \citenamefont {Sugawara},
  \citenamefont {Souma}, \citenamefont {Takahashi},\ and\ \citenamefont
  {Ando}}]{Sato2010TlBiSe2}%
  \BibitemOpen
  \bibfield  {author} {\bibinfo {author} {\bibfnamefont {T.}~\bibnamefont
  {Sato}}, \bibinfo {author} {\bibfnamefont {K.}~\bibnamefont {Segawa}},
  \bibinfo {author} {\bibfnamefont {H.}~\bibnamefont {Guo}}, \bibinfo {author}
  {\bibfnamefont {K.}~\bibnamefont {Sugawara}}, \bibinfo {author}
  {\bibfnamefont {S.}~\bibnamefont {Souma}}, \bibinfo {author} {\bibfnamefont
  {T.}~\bibnamefont {Takahashi}},\ and\ \bibinfo {author} {\bibfnamefont
  {Y.}~\bibnamefont {Ando}},\ }\href
  {https://doi.org/10.1103/PhysRevLett.105.136802} {\bibfield  {journal}
  {\bibinfo  {journal} {Phys. Rev. Lett.}\ }\textbf {\bibinfo {volume} {105}},\
  \bibinfo {pages} {136802} (\bibinfo {year} {2010})}\BibitemShut {NoStop}%
\bibitem [{\citenamefont {Kuroda}\ \emph {et~al.}(2010)\citenamefont {Kuroda}
  \emph {et~al.}}]{Kuroda2010TlBiSe2}%
  \BibitemOpen
  \bibfield  {author} {\bibinfo {author} {\bibfnamefont {K.}~\bibnamefont
  {Kuroda}} \emph {et~al.},\ }\href
  {https://doi.org/10.1103/PhysRevLett.105.146801} {\bibfield  {journal}
  {\bibinfo  {journal} {Phys. Rev. Lett.}\ }\textbf {\bibinfo {volume} {105}},\
  \bibinfo {pages} {146801} (\bibinfo {year} {2010})}\BibitemShut {NoStop}%
\bibitem [{\citenamefont {Chen}\ \emph {et~al.}(2010)\citenamefont {Chen},
  \citenamefont {Liu}, \citenamefont {Analytis}, \citenamefont {Chu},
  \citenamefont {Zhang}, \citenamefont {Yan}, \citenamefont {Mo}, \citenamefont
  {Moore}, \citenamefont {Lu}, \citenamefont {Fisher}, \citenamefont {Zhang},
  \citenamefont {Hussain},\ and\ \citenamefont {Shen}}]{Chen2010TlBiTe2}%
  \BibitemOpen
  \bibfield  {author} {\bibinfo {author} {\bibfnamefont {Y.~L.}\ \bibnamefont
  {Chen}}, \bibinfo {author} {\bibfnamefont {Z.~K.}\ \bibnamefont {Liu}},
  \bibinfo {author} {\bibfnamefont {J.~G.}\ \bibnamefont {Analytis}}, \bibinfo
  {author} {\bibfnamefont {J.-H.}\ \bibnamefont {Chu}}, \bibinfo {author}
  {\bibfnamefont {H.~J.}\ \bibnamefont {Zhang}}, \bibinfo {author}
  {\bibfnamefont {B.~H.}\ \bibnamefont {Yan}}, \bibinfo {author} {\bibfnamefont
  {S.-K.}\ \bibnamefont {Mo}}, \bibinfo {author} {\bibfnamefont {R.~G.}\
  \bibnamefont {Moore}}, \bibinfo {author} {\bibfnamefont {D.~H.}\ \bibnamefont
  {Lu}}, \bibinfo {author} {\bibfnamefont {I.~R.}\ \bibnamefont {Fisher}},
  \bibinfo {author} {\bibfnamefont {S.~C.}\ \bibnamefont {Zhang}}, \bibinfo
  {author} {\bibfnamefont {Z.}~\bibnamefont {Hussain}},\ and\ \bibinfo {author}
  {\bibfnamefont {Z.-X.}\ \bibnamefont {Shen}},\ }\href
  {https://doi.org/10.1103/PhysRevLett.105.266401} {\bibfield  {journal}
  {\bibinfo  {journal} {Phys. Rev. Lett.}\ }\textbf {\bibinfo {volume} {105}},\
  \bibinfo {pages} {266401} (\bibinfo {year} {2010})}\BibitemShut {NoStop}%
\bibitem [{\citenamefont {Hsieh}\ \emph {et~al.}(2009)\citenamefont {Hsieh}
  \emph {et~al.}}]{Hsieh2009Helical}%
  \BibitemOpen
  \bibfield  {author} {\bibinfo {author} {\bibfnamefont {D.}~\bibnamefont
  {Hsieh}} \emph {et~al.},\ }\href {https://doi.org/10.1038/nature08234}
  {\bibfield  {journal} {\bibinfo  {journal} {Nature}\ }\textbf {\bibinfo
  {volume} {460}},\ \bibinfo {pages} {1101} (\bibinfo {year}
  {2009})}\BibitemShut {NoStop}%
\bibitem [{\citenamefont {Gross}\ and\ \citenamefont
  {Neveu}(1974)}]{GrossNeveu1974}%
  \BibitemOpen
  \bibfield  {author} {\bibinfo {author} {\bibfnamefont {D.~J.}\ \bibnamefont
  {Gross}}\ and\ \bibinfo {author} {\bibfnamefont {A.}~\bibnamefont {Neveu}},\
  }\href {https://doi.org/10.1103/PhysRevD.10.3235} {\bibfield  {journal}
  {\bibinfo  {journal} {Phys. Rev. D}\ }\textbf {\bibinfo {volume} {10}},\
  \bibinfo {pages} {3235} (\bibinfo {year} {1974})}\BibitemShut {NoStop}%
\bibitem [{\citenamefont {Herbut}(2006)}]{Herbut2006}%
  \BibitemOpen
  \bibfield  {author} {\bibinfo {author} {\bibfnamefont {I.~F.}\ \bibnamefont
  {Herbut}},\ }\href {https://doi.org/10.1103/PhysRevLett.97.146401} {\bibfield
   {journal} {\bibinfo  {journal} {Phys. Rev. Lett.}\ }\textbf {\bibinfo
  {volume} {97}},\ \bibinfo {pages} {146401} (\bibinfo {year}
  {2006})}\BibitemShut {NoStop}%
\bibitem [{\citenamefont {Rosa}\ \emph {et~al.}(2001)\citenamefont {Rosa},
  \citenamefont {Vitale},\ and\ \citenamefont {Wetterich}}]{Rosa2001FRG}%
  \BibitemOpen
  \bibfield  {author} {\bibinfo {author} {\bibfnamefont {L.}~\bibnamefont
  {Rosa}}, \bibinfo {author} {\bibfnamefont {P.}~\bibnamefont {Vitale}},\ and\
  \bibinfo {author} {\bibfnamefont {C.}~\bibnamefont {Wetterich}},\ }\href
  {https://doi.org/10.1103/PhysRevLett.86.958} {\bibfield  {journal} {\bibinfo
  {journal} {Phys. Rev. Lett.}\ }\textbf {\bibinfo {volume} {86}},\ \bibinfo
  {pages} {958} (\bibinfo {year} {2001})}\BibitemShut {NoStop}%
\bibitem [{\citenamefont {Zerf}\ \emph {et~al.}(2017)\citenamefont {Zerf},
  \citenamefont {Mihaila}, \citenamefont {Marquard}, \citenamefont {Herbut},\
  and\ \citenamefont {Scherer}}]{Zerf2017FourLoop}%
  \BibitemOpen
  \bibfield  {author} {\bibinfo {author} {\bibfnamefont {N.}~\bibnamefont
  {Zerf}}, \bibinfo {author} {\bibfnamefont {L.~N.}\ \bibnamefont {Mihaila}},
  \bibinfo {author} {\bibfnamefont {P.}~\bibnamefont {Marquard}}, \bibinfo
  {author} {\bibfnamefont {I.~F.}\ \bibnamefont {Herbut}},\ and\ \bibinfo
  {author} {\bibfnamefont {M.~M.}\ \bibnamefont {Scherer}},\ }\href
  {https://doi.org/10.1103/PhysRevD.96.096010} {\bibfield  {journal} {\bibinfo
  {journal} {Phys. Rev. D}\ }\textbf {\bibinfo {volume} {96}},\ \bibinfo
  {pages} {096010} (\bibinfo {year} {2017})}\BibitemShut {NoStop}%
\bibitem [{\citenamefont {Liu}\ \emph {et~al.}(2020)\citenamefont {Liu},
  \citenamefont {Wang}, \citenamefont {Sun},\ and\ \citenamefont
  {Meng}}]{Liu2020Designer}%
  \BibitemOpen
  \bibfield  {author} {\bibinfo {author} {\bibfnamefont {Y.}~\bibnamefont
  {Liu}}, \bibinfo {author} {\bibfnamefont {W.}~\bibnamefont {Wang}}, \bibinfo
  {author} {\bibfnamefont {K.}~\bibnamefont {Sun}},\ and\ \bibinfo {author}
  {\bibfnamefont {Z.~Y.}\ \bibnamefont {Meng}},\ }\href
  {https://doi.org/10.1103/PhysRevB.101.064308} {\bibfield  {journal} {\bibinfo
   {journal} {Phys. Rev. B}\ }\textbf {\bibinfo {volume} {101}},\ \bibinfo
  {pages} {064308} (\bibinfo {year} {2020})}\BibitemShut {NoStop}%
\bibitem [{\citenamefont {Wang}\ and\ \citenamefont
  {Meng}(2023)}]{Wang2023EMUS}%
  \BibitemOpen
  \bibfield  {author} {\bibinfo {author} {\bibfnamefont {T.-T.}\ \bibnamefont
  {Wang}}\ and\ \bibinfo {author} {\bibfnamefont {Z.~Y.}\ \bibnamefont
  {Meng}},\ }\href {https://doi.org/10.1103/PhysRevB.108.L121112} {\bibfield
  {journal} {\bibinfo  {journal} {Phys. Rev. B}\ }\textbf {\bibinfo {volume}
  {108}},\ \bibinfo {pages} {L121112} (\bibinfo {year} {2023})}\BibitemShut
  {NoStop}%
\bibitem [{\citenamefont {Tabatabaei}\ \emph {et~al.}(2022)\citenamefont
  {Tabatabaei}, \citenamefont {Negari}, \citenamefont {Maciejko},\ and\
  \citenamefont {Vaezi}}]{Tabatabaei2022}%
  \BibitemOpen
  \bibfield  {author} {\bibinfo {author} {\bibfnamefont {S.~M.}\ \bibnamefont
  {Tabatabaei}}, \bibinfo {author} {\bibfnamefont {A.-R.}\ \bibnamefont
  {Negari}}, \bibinfo {author} {\bibfnamefont {J.}~\bibnamefont {Maciejko}},\
  and\ \bibinfo {author} {\bibfnamefont {A.}~\bibnamefont {Vaezi}},\ }\href
  {https://doi.org/10.1103/PhysRevLett.128.225701} {\bibfield  {journal}
  {\bibinfo  {journal} {Phys. Rev. Lett.}\ }\textbf {\bibinfo {volume} {128}},\
  \bibinfo {pages} {225701} (\bibinfo {year} {2022})}\BibitemShut {NoStop}%
\bibitem [{\citenamefont {Erramilli}\ \emph {et~al.}(2023)\citenamefont
  {Erramilli}, \citenamefont {Iliesiu}, \citenamefont {Kravchuk}, \citenamefont
  {Liu}, \citenamefont {Poland},\ and\ \citenamefont
  {Simmons-Duffin}}]{Erramilli2023}%
  \BibitemOpen
  \bibfield  {author} {\bibinfo {author} {\bibfnamefont {R.~S.}\ \bibnamefont
  {Erramilli}}, \bibinfo {author} {\bibfnamefont {L.~V.}\ \bibnamefont
  {Iliesiu}}, \bibinfo {author} {\bibfnamefont {P.}~\bibnamefont {Kravchuk}},
  \bibinfo {author} {\bibfnamefont {A.}~\bibnamefont {Liu}}, \bibinfo {author}
  {\bibfnamefont {D.}~\bibnamefont {Poland}},\ and\ \bibinfo {author}
  {\bibfnamefont {D.}~\bibnamefont {Simmons-Duffin}},\ }\href
  {https://doi.org/10.1007/JHEP02(2023)036} {\bibfield  {journal} {\bibinfo
  {journal} {J. High Energy Phys.}\ }\textbf {\bibinfo {volume} {2023}}\bibinfo
   {number} { (2)},\ \bibinfo {pages} {036}}\BibitemShut {NoStop}%
\bibitem [{\citenamefont {Nielsen}\ and\ \citenamefont
  {Ninomiya}(1981)}]{NielsenNinomiya1981}%
  \BibitemOpen
\bibfield  {number} {  }\bibfield  {author} {\bibinfo {author} {\bibfnamefont
  {H.~B.}\ \bibnamefont {Nielsen}}\ and\ \bibinfo {author} {\bibfnamefont
  {M.}~\bibnamefont {Ninomiya}},\ }\href
  {https://doi.org/10.1016/0550-3213(81)90361-8} {\bibfield  {journal}
  {\bibinfo  {journal} {Nucl. Phys. B}\ }\textbf {\bibinfo {volume} {185}},\
  \bibinfo {pages} {20} (\bibinfo {year} {1981})}\BibitemShut {NoStop}%
\bibitem [{\citenamefont {Redlich}(1984)}]{Redlich1984}%
  \BibitemOpen
  \bibfield  {author} {\bibinfo {author} {\bibfnamefont {A.~N.}\ \bibnamefont
  {Redlich}},\ }\href {https://doi.org/10.1103/PhysRevLett.52.18} {\bibfield
  {journal} {\bibinfo  {journal} {Phys. Rev. Lett.}\ }\textbf {\bibinfo
  {volume} {52}},\ \bibinfo {pages} {18} (\bibinfo {year} {1984})}\BibitemShut
  {NoStop}%
\bibitem [{\citenamefont {Drell}\ \emph {et~al.}(1976)\citenamefont {Drell},
  \citenamefont {Weinstein},\ and\ \citenamefont
  {Yankielowicz}}]{Drell1976SLAC}%
  \BibitemOpen
  \bibfield  {author} {\bibinfo {author} {\bibfnamefont {S.~D.}\ \bibnamefont
  {Drell}}, \bibinfo {author} {\bibfnamefont {M.}~\bibnamefont {Weinstein}},\
  and\ \bibinfo {author} {\bibfnamefont {S.}~\bibnamefont {Yankielowicz}},\
  }\href {https://doi.org/10.1103/PhysRevD.14.1627} {\bibfield  {journal}
  {\bibinfo  {journal} {Phys. Rev. D}\ }\textbf {\bibinfo {volume} {14}},\
  \bibinfo {pages} {1627} (\bibinfo {year} {1976})}\BibitemShut {NoStop}%
\bibitem [{\citenamefont {Rabin}(1981)}]{rabinPerturbation1981}%
  \BibitemOpen
  \bibfield  {author} {\bibinfo {author} {\bibfnamefont {J.~M.}\ \bibnamefont
  {Rabin}},\ }\href {https://doi.org/10.1103/PhysRevD.24.3218} {\bibfield
  {journal} {\bibinfo  {journal} {Phys. Rev. D}\ }\textbf {\bibinfo {volume}
  {24}},\ \bibinfo {pages} {3218} (\bibinfo {year} {1981})}\BibitemShut
  {NoStop}%
\bibitem [{\citenamefont {Pelissetto}(1988)}]{pelissettoLattice1988}%
  \BibitemOpen
  \bibfield  {author} {\bibinfo {author} {\bibfnamefont {A.}~\bibnamefont
  {Pelissetto}},\ }\href
  {https://doi.org/https://doi.org/10.1016/0003-4916(88)90299-0} {\bibfield
  {journal} {\bibinfo  {journal} {Annals of Physics}\ }\textbf {\bibinfo
  {volume} {182}},\ \bibinfo {pages} {177} (\bibinfo {year}
  {1988})}\BibitemShut {NoStop}%
\bibitem [{\citenamefont {Yu}\ \emph {et~al.}(2026)\citenamefont {Yu},
  \citenamefont {Li}, \citenamefont {Yin},\ and\ \citenamefont
  {Li}}]{Yu2026SA}%
  \BibitemOpen
  \bibfield  {author} {\bibinfo {author} {\bibfnamefont {Y.-K.}\ \bibnamefont
  {Yu}}, \bibinfo {author} {\bibfnamefont {Z.-X.}\ \bibnamefont {Li}}, \bibinfo
  {author} {\bibfnamefont {S.}~\bibnamefont {Yin}},\ and\ \bibinfo {author}
  {\bibfnamefont {Z.-X.}\ \bibnamefont {Li}},\ }\href
  {https://doi.org/10.1126/sciadv.adz4856} {\bibfield  {journal} {\bibinfo
  {journal} {Sci. Adv.}\ }\textbf {\bibinfo {volume} {12}},\ \bibinfo {pages}
  {eadz4856} (\bibinfo {year} {2026})}\BibitemShut {NoStop}%
\bibitem [{\citenamefont {Zhu}\ \emph {et~al.}(2023)\citenamefont {Zhu},
  \citenamefont {Han}, \citenamefont {Huffman}, \citenamefont {Hofmann},\ and\
  \citenamefont {He}}]{Zhu2023FuzzySphere}%
  \BibitemOpen
  \bibfield  {author} {\bibinfo {author} {\bibfnamefont {W.}~\bibnamefont
  {Zhu}}, \bibinfo {author} {\bibfnamefont {C.}~\bibnamefont {Han}}, \bibinfo
  {author} {\bibfnamefont {E.}~\bibnamefont {Huffman}}, \bibinfo {author}
  {\bibfnamefont {J.~S.}\ \bibnamefont {Hofmann}},\ and\ \bibinfo {author}
  {\bibfnamefont {Y.-C.}\ \bibnamefont {He}},\ }\href
  {https://doi.org/10.1103/PhysRevX.13.021009} {\bibfield  {journal} {\bibinfo
  {journal} {Phys. Rev. X}\ }\textbf {\bibinfo {volume} {13}},\ \bibinfo
  {pages} {021009} (\bibinfo {year} {2023})}\BibitemShut {NoStop}%
\bibitem [{\citenamefont {Hu}\ \emph {et~al.}(2023)\citenamefont {Hu},
  \citenamefont {He},\ and\ \citenamefont {Zhu}}]{Hu2023SphereOPE}%
  \BibitemOpen
  \bibfield  {author} {\bibinfo {author} {\bibfnamefont {L.}~\bibnamefont
  {Hu}}, \bibinfo {author} {\bibfnamefont {Y.-C.}\ \bibnamefont {He}},\ and\
  \bibinfo {author} {\bibfnamefont {W.}~\bibnamefont {Zhu}},\ }\href
  {https://doi.org/10.1103/PhysRevLett.131.031601} {\bibfield  {journal}
  {\bibinfo  {journal} {Phys. Rev. Lett.}\ }\textbf {\bibinfo {volume} {131}},\
  \bibinfo {pages} {031601} (\bibinfo {year} {2023})}\BibitemShut {NoStop}%
\bibitem [{\citenamefont {Zhou}\ \emph {et~al.}(2024)\citenamefont {Zhou},
  \citenamefont {Hu}, \citenamefont {Zhu},\ and\ \citenamefont
  {He}}]{Zhou2024SphereSO5}%
  \BibitemOpen
  \bibfield  {author} {\bibinfo {author} {\bibfnamefont {Z.}~\bibnamefont
  {Zhou}}, \bibinfo {author} {\bibfnamefont {L.}~\bibnamefont {Hu}}, \bibinfo
  {author} {\bibfnamefont {W.}~\bibnamefont {Zhu}},\ and\ \bibinfo {author}
  {\bibfnamefont {Y.-C.}\ \bibnamefont {He}},\ }\href
  {https://doi.org/10.1103/PhysRevX.14.021044} {\bibfield  {journal} {\bibinfo
  {journal} {Phys. Rev. X}\ }\textbf {\bibinfo {volume} {14}},\ \bibinfo
  {pages} {021044} (\bibinfo {year} {2024})}\BibitemShut {NoStop}%
\bibitem [{\citenamefont {Chen}\ \emph
  {et~al.}(2024{\natexlab{a}})\citenamefont {Chen}, \citenamefont {Zhang},
  \citenamefont {Wang}, \citenamefont {Sun},\ and\ \citenamefont
  {Meng}}]{ChenPhase2024}%
  \BibitemOpen
  \bibfield  {author} {\bibinfo {author} {\bibfnamefont {B.-B.}\ \bibnamefont
  {Chen}}, \bibinfo {author} {\bibfnamefont {X.}~\bibnamefont {Zhang}},
  \bibinfo {author} {\bibfnamefont {Y.}~\bibnamefont {Wang}}, \bibinfo {author}
  {\bibfnamefont {K.}~\bibnamefont {Sun}},\ and\ \bibinfo {author}
  {\bibfnamefont {Z.~Y.}\ \bibnamefont {Meng}},\ }\href
  {https://doi.org/10.1103/PhysRevLett.132.246503} {\bibfield  {journal}
  {\bibinfo  {journal} {Phys. Rev. Lett.}\ }\textbf {\bibinfo {volume} {132}},\
  \bibinfo {pages} {246503} (\bibinfo {year} {2024}{\natexlab{a}})}\BibitemShut
  {NoStop}%
\bibitem [{\citenamefont {Chen}\ \emph
  {et~al.}(2024{\natexlab{b}})\citenamefont {Chen}, \citenamefont {Zhang},\
  and\ \citenamefont {Meng}}]{chenEmergent2024}%
  \BibitemOpen
  \bibfield  {author} {\bibinfo {author} {\bibfnamefont {B.-B.}\ \bibnamefont
  {Chen}}, \bibinfo {author} {\bibfnamefont {X.}~\bibnamefont {Zhang}},\ and\
  \bibinfo {author} {\bibfnamefont {Z.~Y.}\ \bibnamefont {Meng}},\ }\href
  {https://doi.org/10.1103/PhysRevB.110.125153} {\bibfield  {journal} {\bibinfo
   {journal} {Phys. Rev. B}\ }\textbf {\bibinfo {volume} {110}},\ \bibinfo
  {pages} {125153} (\bibinfo {year} {2024}{\natexlab{b}})}\BibitemShut
  {NoStop}%
\bibitem [{\citenamefont {Liao}\ \emph {et~al.}(2026)\citenamefont {Liao},
  \citenamefont {Chen}, \citenamefont {Rong}, \citenamefont {Assaad},
  \citenamefont {Janssen},\ and\ \citenamefont {Meng}}]{Liao2026SO5WZW}%
  \BibitemOpen
  \bibfield  {author} {\bibinfo {author} {\bibfnamefont {Y.~D.}\ \bibnamefont
  {Liao}}, \bibinfo {author} {\bibfnamefont {B.-B.}\ \bibnamefont {Chen}},
  \bibinfo {author} {\bibfnamefont {J.}~\bibnamefont {Rong}}, \bibinfo {author}
  {\bibfnamefont {F.~F.}\ \bibnamefont {Assaad}}, \bibinfo {author}
  {\bibfnamefont {L.}~\bibnamefont {Janssen}},\ and\ \bibinfo {author}
  {\bibfnamefont {Z.~Y.}\ \bibnamefont {Meng}},\ }\href@noop {} {\bibinfo
  {title} {Numerical evidence of a critical point in the (2+1)d {SO(5)}
  nonlinear sigma model with {Wess-Zumino-Witten} term}} (\bibinfo {year}
  {2026}),\ \Eprint {https://arxiv.org/abs/2605.03700} {arXiv:2605.03700}
  \BibitemShut {NoStop}%
\bibitem [{\citenamefont {Huffman}\ \emph {et~al.}(2026)\citenamefont
  {Huffman}, \citenamefont {Zhou}, \citenamefont {He},\ and\ \citenamefont
  {Hofmann}}]{Huffman2026SphereQCD}%
  \BibitemOpen
  \bibfield  {author} {\bibinfo {author} {\bibfnamefont {E.}~\bibnamefont
  {Huffman}}, \bibinfo {author} {\bibfnamefont {Z.}~\bibnamefont {Zhou}},
  \bibinfo {author} {\bibfnamefont {Y.-C.}\ \bibnamefont {He}},\ and\ \bibinfo
  {author} {\bibfnamefont {J.~S.}\ \bibnamefont {Hofmann}},\ }\href
  {https://doi.org/10.1103/f6qg-q875} {\bibfield  {journal} {\bibinfo
  {journal} {Phys. Rev. D}\ }\textbf {\bibinfo {volume} {114}},\ \bibinfo
  {pages} {034503} (\bibinfo {year} {2026})}\BibitemShut {NoStop}%
\bibitem [{\citenamefont {Gao}\ \emph {et~al.}(2026)\citenamefont {Gao},
  \citenamefont {Wang},\ and\ \citenamefont {Lee}}]{Gao2026Sphere}%
  \BibitemOpen
  \bibfield  {author} {\bibinfo {author} {\bibfnamefont {Z.-Q.}\ \bibnamefont
  {Gao}}, \bibinfo {author} {\bibfnamefont {T.}~\bibnamefont {Wang}},\ and\
  \bibinfo {author} {\bibfnamefont {D.-H.}\ \bibnamefont {Lee}},\ }\href
  {https://doi.org/10.1103/b8b7-scc5} {\bibfield  {journal} {\bibinfo
  {journal} {Phys. Rev. Lett.}\ }\textbf {\bibinfo {volume} {137}},\ \bibinfo
  {pages} {166603} (\bibinfo {year} {2026})}\BibitemShut {NoStop}%
\bibitem [{Sup()}]{SupplementalMaterial}%
  \BibitemOpen
  \href@noop {} {}\bibinfo {note} {See Supplemental Material for orbital matrix
  elements, the centered imaginary-density decoupling and layer updates, phase
  reweighting, observable estimators, the time-step comparison with exact
  diagonalization, and finite-size fits}\BibitemShut {NoStop}%
\bibitem [{\citenamefont {Troyer}\ and\ \citenamefont
  {Wiese}(2005)}]{Troyer2005}%
  \BibitemOpen
  \bibfield  {author} {\bibinfo {author} {\bibfnamefont {M.}~\bibnamefont
  {Troyer}}\ and\ \bibinfo {author} {\bibfnamefont {U.-J.}\ \bibnamefont
  {Wiese}},\ }\href {https://doi.org/10.1103/PhysRevLett.94.170201} {\bibfield
  {journal} {\bibinfo  {journal} {Phys. Rev. Lett.}\ }\textbf {\bibinfo
  {volume} {94}},\ \bibinfo {pages} {170201} (\bibinfo {year}
  {2005})}\BibitemShut {NoStop}%
\bibitem [{\citenamefont {Pan}\ and\ \citenamefont {Meng}(2024)}]{panSign2024}%
  \BibitemOpen
  \bibfield  {author} {\bibinfo {author} {\bibfnamefont {G.}~\bibnamefont
  {Pan}}\ and\ \bibinfo {author} {\bibfnamefont {Z.~Y.}\ \bibnamefont {Meng}},\
  }in\ \href
  {https://doi.org/https://doi.org/10.1016/B978-0-323-90800-9.00095-0} {\emph
  {\bibinfo {booktitle} {Encyclopedia of Condensed Matter Physics (Second
  Edition)}}},\ \bibinfo {editor} {edited by\ \bibinfo {editor} {\bibfnamefont
  {T.}~\bibnamefont {Chakraborty}}}\ (\bibinfo  {publisher} {Academic Press},\
  \bibinfo {address} {Oxford},\ \bibinfo {year} {2024})\ \bibinfo {edition}
  {second edition}\ ed.,\ pp.\ \bibinfo {pages} {879--893}\BibitemShut
  {NoStop}%
\bibitem [{\citenamefont {Huang}\ \emph {et~al.}(2025)\citenamefont {Huang},
  \citenamefont {Parthenios}, \citenamefont {Ulybyshev}, \citenamefont {Zhang},
  \citenamefont {Assaad}, \citenamefont {Classen},\ and\ \citenamefont
  {Meng}}]{Huang2025AngleGN}%
  \BibitemOpen
  \bibfield  {author} {\bibinfo {author} {\bibfnamefont {C.}~\bibnamefont
  {Huang}}, \bibinfo {author} {\bibfnamefont {N.}~\bibnamefont {Parthenios}},
  \bibinfo {author} {\bibfnamefont {M.}~\bibnamefont {Ulybyshev}}, \bibinfo
  {author} {\bibfnamefont {X.}~\bibnamefont {Zhang}}, \bibinfo {author}
  {\bibfnamefont {F.~F.}\ \bibnamefont {Assaad}}, \bibinfo {author}
  {\bibfnamefont {L.}~\bibnamefont {Classen}},\ and\ \bibinfo {author}
  {\bibfnamefont {Z.~Y.}\ \bibnamefont {Meng}},\ }\href
  {https://doi.org/10.1038/s41467-025-62461-y} {\bibfield  {journal} {\bibinfo
  {journal} {Nature Communications}\ }\textbf {\bibinfo {volume} {16}},\
  \bibinfo {pages} {7176} (\bibinfo {year} {2025})}\BibitemShut {NoStop}%
\bibitem [{\citenamefont {{Shan}}\ \emph
  {et~al.}(2026{\natexlab{a}})\citenamefont {{Shan}}, \citenamefont {{Huang}},
  \citenamefont {{Ledwith}},\ and\ \citenamefont {{Meng}}}]{shanTrion2026}%
  \BibitemOpen
  \bibfield  {author} {\bibinfo {author} {\bibfnamefont {S.}~\bibnamefont
  {{Shan}}}, \bibinfo {author} {\bibfnamefont {C.}~\bibnamefont {{Huang}}},
  \bibinfo {author} {\bibfnamefont {P.}~\bibnamefont {{Ledwith}}},\ and\
  \bibinfo {author} {\bibfnamefont {Z.~Y.}\ \bibnamefont {{Meng}}},\
  }\href@noop {} {\bibinfo {title} {{Trion Excitations in Twisted Bilayer
  Graphene: A Quantum Monte Carlo Study}}} (\bibinfo {year}
  {2026}{\natexlab{a}}),\ \Eprint {https://arxiv.org/abs/2607.27745}
  {arXiv:2607.27745} \BibitemShut {NoStop}%
\bibitem [{\citenamefont {Shi}\ \emph {et~al.}(2015)\citenamefont {Shi},
  \citenamefont {Chiesa},\ and\ \citenamefont {Zhang}}]{Shi2015ForceBias}%
  \BibitemOpen
  \bibfield  {author} {\bibinfo {author} {\bibfnamefont {H.}~\bibnamefont
  {Shi}}, \bibinfo {author} {\bibfnamefont {S.}~\bibnamefont {Chiesa}},\ and\
  \bibinfo {author} {\bibfnamefont {S.}~\bibnamefont {Zhang}},\ }\href
  {https://doi.org/10.1103/PhysRevA.92.033603} {\bibfield  {journal} {\bibinfo
  {journal} {Phys. Rev. A}\ }\textbf {\bibinfo {volume} {92}},\ \bibinfo
  {pages} {033603} (\bibinfo {year} {2015})}\BibitemShut {NoStop}%
\bibitem [{\citenamefont {Hastings}(1970)}]{Hastings1970}%
  \BibitemOpen
  \bibfield  {author} {\bibinfo {author} {\bibfnamefont {W.~K.}\ \bibnamefont
  {Hastings}},\ }\href {https://doi.org/10.1093/biomet/57.1.97} {\bibfield
  {journal} {\bibinfo  {journal} {Biometrika}\ }\textbf {\bibinfo {volume}
  {57}},\ \bibinfo {pages} {97} (\bibinfo {year} {1970})}\BibitemShut {NoStop}%
\bibitem [{\citenamefont {Qin}\ \emph {et~al.}(2017)\citenamefont {Qin},
  \citenamefont {He}, \citenamefont {You}, \citenamefont {Lu}, \citenamefont
  {Sen}, \citenamefont {Sandvik}, \citenamefont {Xu},\ and\ \citenamefont
  {Meng}}]{qinDuality2017}%
  \BibitemOpen
  \bibfield  {author} {\bibinfo {author} {\bibfnamefont {Y.~Q.}\ \bibnamefont
  {Qin}}, \bibinfo {author} {\bibfnamefont {Y.-Y.}\ \bibnamefont {He}},
  \bibinfo {author} {\bibfnamefont {Y.-Z.}\ \bibnamefont {You}}, \bibinfo
  {author} {\bibfnamefont {Z.-Y.}\ \bibnamefont {Lu}}, \bibinfo {author}
  {\bibfnamefont {A.}~\bibnamefont {Sen}}, \bibinfo {author} {\bibfnamefont
  {A.~W.}\ \bibnamefont {Sandvik}}, \bibinfo {author} {\bibfnamefont
  {C.}~\bibnamefont {Xu}},\ and\ \bibinfo {author} {\bibfnamefont {Z.~Y.}\
  \bibnamefont {Meng}},\ }\href {https://doi.org/10.1103/PhysRevX.7.031052}
  {\bibfield  {journal} {\bibinfo  {journal} {Phys. Rev. X}\ }\textbf {\bibinfo
  {volume} {7}},\ \bibinfo {pages} {031052} (\bibinfo {year}
  {2017})}\BibitemShut {NoStop}%
\bibitem [{\citenamefont {Iliesiu}\ \emph {et~al.}(2018)\citenamefont
  {Iliesiu}, \citenamefont {Kos}, \citenamefont {Poland}, \citenamefont
  {Pufu},\ and\ \citenamefont {Simmons-Duffin}}]{Iliesiu2018}%
  \BibitemOpen
  \bibfield  {author} {\bibinfo {author} {\bibfnamefont {L.}~\bibnamefont
  {Iliesiu}}, \bibinfo {author} {\bibfnamefont {F.}~\bibnamefont {Kos}},
  \bibinfo {author} {\bibfnamefont {D.}~\bibnamefont {Poland}}, \bibinfo
  {author} {\bibfnamefont {S.~S.}\ \bibnamefont {Pufu}},\ and\ \bibinfo
  {author} {\bibfnamefont {D.}~\bibnamefont {Simmons-Duffin}},\ }\href
  {https://doi.org/10.1007/JHEP01(2018)036} {\bibfield  {journal} {\bibinfo
  {journal} {J. High Energy Phys.}\ }\textbf {\bibinfo {volume} {2018}}\bibinfo
   {number} { (1)},\ \bibinfo {pages} {036}}\BibitemShut {NoStop}%
\bibitem [{\citenamefont {{Shan}}\ \emph
  {et~al.}(2026{\natexlab{b}})\citenamefont {{Shan}}, \citenamefont {{Huang}},\
  and\ \citenamefont {{Meng}}}]{shanMomentum2026}%
  \BibitemOpen
\bibfield  {number} {  }\bibfield  {author} {\bibinfo {author} {\bibfnamefont
  {S.}~\bibnamefont {{Shan}}}, \bibinfo {author} {\bibfnamefont
  {C.}~\bibnamefont {{Huang}}},\ and\ \bibinfo {author} {\bibfnamefont {Z.~Y.}\
  \bibnamefont {{Meng}}},\ }\href {https://doi.org/10.48550/arXiv.2611.25522}
  {\bibfield  {journal} {\bibinfo  {journal} {arXiv e-prints}\ ,\ \bibinfo
  {eid} {arXiv:2611.25522}} (\bibinfo {year} {2026}{\natexlab{b}})},\ \Eprint
  {https://arxiv.org/abs/2611.25522} {arXiv:2611.25522 [cond-mat.str-el]}
  \BibitemShut {NoStop}%
\bibitem [{\citenamefont {Hubbard}(1959)}]{Hubbard1959HS}%
  \BibitemOpen
  \bibfield  {author} {\bibinfo {author} {\bibfnamefont {J.}~\bibnamefont
  {Hubbard}},\ }\href {https://doi.org/10.1103/PhysRevLett.3.77} {\bibfield
  {journal} {\bibinfo  {journal} {Phys. Rev. Lett.}\ }\textbf {\bibinfo
  {volume} {3}},\ \bibinfo {pages} {77} (\bibinfo {year} {1959})}\BibitemShut
  {NoStop}%
\bibitem [{\citenamefont {Blankenbecler}\ \emph {et~al.}(1981)\citenamefont
  {Blankenbecler}, \citenamefont {Scalapino},\ and\ \citenamefont
  {Sugar}}]{Blankenbecler1981}%
  \BibitemOpen
  \bibfield  {author} {\bibinfo {author} {\bibfnamefont {R.}~\bibnamefont
  {Blankenbecler}}, \bibinfo {author} {\bibfnamefont {D.~J.}\ \bibnamefont
  {Scalapino}},\ and\ \bibinfo {author} {\bibfnamefont {R.~L.}\ \bibnamefont
  {Sugar}},\ }\href {https://doi.org/10.1103/PhysRevD.24.2278} {\bibfield
  {journal} {\bibinfo  {journal} {Phys. Rev. D}\ }\textbf {\bibinfo {volume}
  {24}},\ \bibinfo {pages} {2278} (\bibinfo {year} {1981})}\BibitemShut
  {NoStop}%
\bibitem [{\citenamefont {Sugiyama}\ and\ \citenamefont
  {Koonin}(1986)}]{Sugiyama1986PQMC}%
  \BibitemOpen
  \bibfield  {author} {\bibinfo {author} {\bibfnamefont {G.}~\bibnamefont
  {Sugiyama}}\ and\ \bibinfo {author} {\bibfnamefont {S.~E.}\ \bibnamefont
  {Koonin}},\ }\href {https://doi.org/10.1016/0003-4916(86)90107-7} {\bibfield
  {journal} {\bibinfo  {journal} {Ann. Phys. (N.Y.)}\ }\textbf {\bibinfo
  {volume} {168}},\ \bibinfo {pages} {1} (\bibinfo {year} {1986})}\BibitemShut
  {NoStop}%
\bibitem [{\citenamefont {Sorella}\ \emph {et~al.}(1989)\citenamefont
  {Sorella}, \citenamefont {Baroni}, \citenamefont {Car},\ and\ \citenamefont
  {Parrinello}}]{Sorella1989PQMC}%
  \BibitemOpen
  \bibfield  {author} {\bibinfo {author} {\bibfnamefont {S.}~\bibnamefont
  {Sorella}}, \bibinfo {author} {\bibfnamefont {S.}~\bibnamefont {Baroni}},
  \bibinfo {author} {\bibfnamefont {R.}~\bibnamefont {Car}},\ and\ \bibinfo
  {author} {\bibfnamefont {M.}~\bibnamefont {Parrinello}},\ }\href
  {https://doi.org/10.1209/0295-5075/8/7/014} {\bibfield  {journal} {\bibinfo
  {journal} {Europhys. Lett.}\ }\textbf {\bibinfo {volume} {8}},\ \bibinfo
  {pages} {663} (\bibinfo {year} {1989})}\BibitemShut {NoStop}%
\bibitem [{\citenamefont {Assaad}\ and\ \citenamefont
  {Evertz}(2008)}]{AssaadEvertz2008}%
  \BibitemOpen
  \bibfield  {author} {\bibinfo {author} {\bibfnamefont {F.~F.}\ \bibnamefont
  {Assaad}}\ and\ \bibinfo {author} {\bibfnamefont {H.~G.}\ \bibnamefont
  {Evertz}},\ }in\ \href {https://doi.org/10.1007/978-3-540-74686-7_10} {\emph
  {\bibinfo {booktitle} {Computational Many-Particle Physics}}},\ \bibinfo
  {series} {Lecture Notes in Physics}, Vol.\ \bibinfo {volume} {739},\ \bibinfo
  {editor} {edited by\ \bibinfo {editor} {\bibfnamefont {H.}~\bibnamefont
  {Fehske}}, \bibinfo {editor} {\bibfnamefont {R.}~\bibnamefont {Schneider}},\
  and\ \bibinfo {editor} {\bibfnamefont {A.}~\bibnamefont {Wei{\ss}e}}}\
  (\bibinfo  {publisher} {Springer},\ \bibinfo {address} {Berlin, Heidelberg},\
  \bibinfo {year} {2008})\ pp.\ \bibinfo {pages} {277--356}\BibitemShut
  {NoStop}%
\bibitem [{\citenamefont {White}\ \emph {et~al.}(1989)\citenamefont {White},
  \citenamefont {Scalapino}, \citenamefont {Sugar}, \citenamefont {Loh},
  \citenamefont {Gubernatis},\ and\ \citenamefont
  {Scalettar}}]{White1989Stability}%
  \BibitemOpen
  \bibfield  {author} {\bibinfo {author} {\bibfnamefont {S.~R.}\ \bibnamefont
  {White}}, \bibinfo {author} {\bibfnamefont {D.~J.}\ \bibnamefont
  {Scalapino}}, \bibinfo {author} {\bibfnamefont {R.~L.}\ \bibnamefont
  {Sugar}}, \bibinfo {author} {\bibfnamefont {E.~Y.}\ \bibnamefont {Loh}},
  \bibinfo {author} {\bibfnamefont {J.~E.}\ \bibnamefont {Gubernatis}},\ and\
  \bibinfo {author} {\bibfnamefont {R.~T.}\ \bibnamefont {Scalettar}},\ }\href
  {https://doi.org/10.1103/PhysRevB.40.506} {\bibfield  {journal} {\bibinfo
  {journal} {Phys. Rev. B}\ }\textbf {\bibinfo {volume} {40}},\ \bibinfo
  {pages} {506} (\bibinfo {year} {1989})}\BibitemShut {NoStop}%
\bibitem [{\citenamefont {Roberts}\ and\ \citenamefont
  {Rosenthal}(1998)}]{Roberts1998MALA}%
  \BibitemOpen
  \bibfield  {author} {\bibinfo {author} {\bibfnamefont {G.~O.}\ \bibnamefont
  {Roberts}}\ and\ \bibinfo {author} {\bibfnamefont {J.~S.}\ \bibnamefont
  {Rosenthal}},\ }\href {https://doi.org/10.1111/1467-9868.00123} {\bibfield
  {journal} {\bibinfo  {journal} {J. R. Stat. Soc. B}\ }\textbf {\bibinfo
  {volume} {60}},\ \bibinfo {pages} {255} (\bibinfo {year} {1998})}\BibitemShut
  {NoStop}%
\bibitem [{\citenamefont {Loh}\ \emph {et~al.}(1990)\citenamefont {Loh},
  \citenamefont {Gubernatis}, \citenamefont {Scalettar}, \citenamefont {White},
  \citenamefont {Scalapino},\ and\ \citenamefont {Sugar}}]{Loh1990Sign}%
  \BibitemOpen
  \bibfield  {author} {\bibinfo {author} {\bibfnamefont {E.~Y.}\ \bibnamefont
  {Loh}, \bibfnamefont {Jr.}}, \bibinfo {author} {\bibfnamefont {J.~E.}\
  \bibnamefont {Gubernatis}}, \bibinfo {author} {\bibfnamefont {R.~T.}\
  \bibnamefont {Scalettar}}, \bibinfo {author} {\bibfnamefont {S.~R.}\
  \bibnamefont {White}}, \bibinfo {author} {\bibfnamefont {D.~J.}\ \bibnamefont
  {Scalapino}},\ and\ \bibinfo {author} {\bibfnamefont {R.~L.}\ \bibnamefont
  {Sugar}},\ }\href {https://doi.org/10.1103/PhysRevB.41.9301} {\bibfield
  {journal} {\bibinfo  {journal} {Phys. Rev. B}\ }\textbf {\bibinfo {volume}
  {41}},\ \bibinfo {pages} {9301} (\bibinfo {year} {1990})}\BibitemShut
  {NoStop}%
\end{thebibliography}%
\clearpage
\onecolumngrid
\appendix
\setcounter{figure}{0}
\setcounter{table}{0}
\setcounter{equation}{0}
\renewcommand{\theequation}{S\arabic{equation}}
\renewcommand{\theHequation}{supp.\arabic{equation}}
\renewcommand{\thefigure}{S\arabic{figure}}
\renewcommand{\thetable}{S\arabic{table}}
\renewcommand{\theHfigure}{supp.\arabic{figure}}
\renewcommand{\theHtable}{supp.\arabic{table}}
\section{Supplemental Material: projector and analysis details}

\subsection{Orbital basis and interaction vertices}
\label{sec:sm-basis}
Write $I=(s,k,m)$, where $s=\pm$, $k=1,\ldots,n_{\rm cut}$,
$j=k-1/2$, and $m=-j,\ldots,j$. The branch ordering is all positive-energy
orbitals followed by all negative-energy orbitals; within each branch $k$
and then $m$ increase. Each branch contains $N_0=n_{\rm cut}(n_{\rm cut}+1)$
orbitals, the one-particle dimension is $D=2N_0$, and the neutral sector has
$N=N_0$. We use $v=1$ and $R=\sqrt{N_0/2}$.
Let $Y^{\rm c}_{\ell m}$ be the usual complex harmonics normalized in
$d\Omega$, with the Condon--Shortley phase and
$Y^{\rm c}_{\ell,-m}=(-1)^m(Y^{\rm c}_{\ell m})^*$.
An explicit gauge for the spinors of the spherical regulator~\cite{Gao2026Sphere} is
\begin{equation}
 u_{+km}(\Omega)=
 \begin{pmatrix}
 \sqrt{\dfrac{j+m}{2j}}\,Y^{\rm c}_{k-1,m-1/2}(\Omega)\\[2pt]
 \sqrt{\dfrac{j-m}{2j}}\,Y^{\rm c}_{k-1,m+1/2}(\Omega)
 \end{pmatrix},\qquad
 u_{-km}(\Omega)=(\hat{\bm n}\cdot\bm\sigma)u_{+km}(\Omega).
 \label{eq:sm-spinors}
\end{equation}
A harmonic with an inadmissible magnetic index is zero. These conventions
fix the relative phase between the two branches. They obey
$(\bm\sigma\cdot\bm L+1)u_{skm}=sk\,u_{skm}$, and their orbital angular
momenta are $k-1$ and $k$ for $s=+$ and $-$, respectively. Thus
\begin{equation}
 K_{IJ}=\frac{s_Ivk_I}{R}\delta_{IJ},\qquad
 \Gamma=\begin{pmatrix}0&I_{N_0}\\I_{N_0}&0\end{pmatrix},\qquad
 \int d\Omega\,m_\Omega=f^\dagger\Gamma f .
 \label{eq:sm-kinetic-mass}
\end{equation}
There is no explicit bulk mass or chemical potential. The trial occupied
columns may be chosen as $Q_T=(0,I_{N_0})^{\rm T}$. In this free sea,
$\langle M\rangle_T=0$ and $\langle M^2\rangle_T=N_0$.

For the interaction channels we use real harmonics $Y_q$, with
$q=(\ell,\mu)$ and
\begin{equation}
 Y_{\ell\mu}=
 \begin{cases}
 \sqrt{2}\,\operatorname{Im}Y^{\rm c}_{\ell,|\mu|},&\mu<0,\\
 Y^{\rm c}_{\ell0},&\mu=0,\\
 \sqrt{2}\,\operatorname{Re}Y^{\rm c}_{\ell\mu},&\mu>0.
 \end{cases}
 \qquad
 \int d\Omega\,Y_qY_{q'}=\delta_{qq'} .
 \label{eq:sm-real-harmonics}
\end{equation}
The mass and charge vertices are the one-particle matrices
\begin{align}
 (A_q)_{IJ}&=\int d\Omega\,Y_q(\Omega)
 u_I^\dagger(\Omega)(\hat{\bm n}\cdot\bm\sigma)u_J(\Omega),\\
 (T_q)_{IJ}&=\int d\Omega\,Y_q(\Omega)
 u_I^\dagger(\Omega)u_J(\Omega).
 \label{eq:sm-matrix-elements}
\end{align}
To evaluate the angular integral, write the spin component with spin projection $\xi=\pm1/2$ as
\begin{equation}
 [u_{skm}]_\xi=b_{skm,\xi}\,Y^{\rm c}_{L_s(k),m-\xi},\qquad
 L_+(k)=k-1,\quad L_-(k)=k .
\end{equation}
Here the coefficients in the gauge of Eq.~\eqref{eq:sm-spinors} are
\begin{align}
 (b_{+,1/2},b_{+,-1/2})&=
 \left(\sqrt{\frac{j+m}{2j}},\sqrt{\frac{j-m}{2j}}\right),\\
 (b_{-,1/2},b_{-,-1/2})&=
 \left(\sqrt{\frac{j-m+1}{2j+2}},-\sqrt{\frac{j+m+1}{2j+2}}\right),
\end{align}
where the common indices $k,m$ are suppressed. Define the coefficients $a_{\mu M}$ by $Y_{\ell\mu}=\sum_{M=-\ell}^{\ell}a_{\mu M}Y^{\rm c}_{\ell M}$ using Eq.~\eqref{eq:sm-real-harmonics}. Only $M=0$ contributes for $\mu=0$, and only $M=\pm|\mu|$ can contribute otherwise. With $L=L_s(k)$, $L'=L_{-s'}(k')$, $p=m-\xi$, and $p'=m'-\xi$, the evaluated vertex is
\begin{align}
 (A_{\ell\mu})_{skm,s'k'm'} &=
 \sum_{\xi,M} b^*_{skm,\xi} b_{-s'k'm',\xi} a_{\mu M}
 (-1)^p
 \sqrt{\frac{(2L+1)(2\ell+1)(2L'+1)}{4\pi}}
 \nonumber\\
 &\quad\times
 \begin{pmatrix}L&\ell&L'\\0&0&0\end{pmatrix}
 \begin{pmatrix}L&\ell&L'\\-p&M&p'\end{pmatrix}.
 \label{eq:sm-gaunt-vertex}
\end{align}
The two $3j$ symbols enforce the orbital triangle rule, even $L+\ell+L'$, and $M=m-m'$. For the charge vertex the radial spin operator is absent. Taking $L'=L_{s'}(k')$ and retaining the other definitions above gives
\begin{align}
 (T_{\ell\mu})_{skm,s'k'm'} &=
 \sum_{\xi,M} b^*_{skm,\xi} b_{s'k'm',\xi} a_{\mu M}
 (-1)^p
 \sqrt{\frac{(2L+1)(2\ell+1)(2L'+1)}{4\pi}}
 \nonumber\\
 &\quad\times
 \begin{pmatrix}L&\ell&L'\\0&0&0\end{pmatrix}
 \begin{pmatrix}L&\ell&L'\\-p&M&p'\end{pmatrix}.
 \label{eq:sm-gaunt-density}
\end{align}
Both are Hermitian. With $Q_q=f^\dagger A_qf$ and $P_q=f^\dagger T_qf$,
$m_\Omega=\sum_qY_q(\Omega)Q_q$ and
$\rho_\Omega=\sum_qY_q(\Omega)P_q$. The angular cutoff is
$0\le\ell\le2n_{\rm cut}-1$, containing $N_q=(2n_{\rm cut})^2$
real channels. The branch gauge in Eq.~\eqref{eq:sm-spinors} gives
\begin{equation}
 T_q=\Gamma A_q=A_q\Gamma,\qquad
 A_{00}=\frac{\Gamma}{\sqrt{4\pi}},\qquad
 T_{00}=\frac{I_D}{\sqrt{4\pi}} .
 \label{eq:sm-vertex-relation}
\end{equation}
In particular, a density sampling vertex is not the physical mass vertex
used to measure $S_\ell$.

Define $b_+=0$ and $b_-=1$. The analytically allowed mass entries satisfy
\begin{equation}
 \begin{gathered}
 |k_I-k_J|\le\ell\le k_I+k_J-1,\qquad
 |m_I-m_J|=|\mu|,\\
 k_I+k_J+b_{s_I}+b_{s_J}+\ell\ \text{odd}.
 \end{gathered}
 \label{eq:sm-mass-selection}
\end{equation}
The density rule has the same triangle and magnetic conditions but even
parity. The magnetic condition accounts for both signs present in a real
harmonic. These are necessary conditions; allowed entries can also vanish.
For each orbital pair, $|\mu|$ is fixed and at most $O(n_{\rm cut})$
channels survive. We retain every matrix element allowed by the angular-momentum selection rules, irrespective of magnitude.

\subsection{Local operator identity and the centered density channels}
\label{sec:density-hs}
Let \(U(\Omega)\) be the \(2\times D\) matrix whose columns
are the spinors defined above.  The complete positive/negative multiplets
obey
\begin{equation}
 U(\Omega)U(\Omega)^\dagger=\kappa I_2,
 \qquad \kappa=\frac{N_0}{4\pi},\qquad
 \{\Psi_a(\Omega),\Psi_b^\dagger(\Omega)\}
 =\kappa\delta_{ab}.
 \label{eq:local-car}
\end{equation}
Substitution of the spinors and the harmonic addition theorem gives
\(\sum_m u_{skm}(\Omega)u_{skm}(\Omega)^\dagger
=kI_2/(4\pi)\) for either branch.  Summing over \(s\) and \(k\) gives
Eq.~\eqref{eq:local-car}.  At a fixed \(\Omega\), diagonalize
\(\hat{\boldsymbol n}\cdot\boldsymbol\sigma\) and divide the two local
field components by \(\sqrt\kappa\).  The resulting operators \(c_\pm\)
have canonical anticommutators.  With \(n_\pm=c_\pm^\dagger c_\pm\),
\(n_\pm^2=n_\pm\) and \([n_+,n_-]=0\), while
\(m_\Omega=\kappa(n_+-n_-)\) and
\(\rho_\Omega=\kappa(n_++n_-)\).  Hence the following is an exact
operator identity, not a continuum Fierz approximation:
\begin{equation}
 m_\Omega^2=\kappa^2-(\rho_\Omega-\kappa)^2,
 \qquad \rho_\Omega=\Psi^\dagger(\Omega)\Psi(\Omega).
 \label{eq:density-fierz}
\end{equation}

Here $m_\Omega^2$ denotes the ordinary operator product.

For the density vertices of Eq.~\eqref{eq:sm-matrix-elements}, define the neutral-sector versions
\begin{equation}
  \bar T_q=T_q-\frac{\operatorname{tr}T_q}{D}I_D,
 \qquad \bar P_q=f^\dagger\bar T_q f .
 \label{eq:centered-vertices}
\end{equation}
Here and below \(Y_q\) denotes the real harmonic with the convention given
above.  Since \(\operatorname{tr}T_q=2\kappa\int d\Omega\,Y_q\) and
\(\hat N=N_0=D/2\), the trace subtraction is exactly the scalar centering:
\begin{equation}
 \bar P_q=\int d\Omega\,Y_q(\Omega)(\rho_\Omega-\kappa),
 \qquad
 H=f^\dagger Kf+\frac{u_0}{2}\sum_q\bar P_q^2-C_0,
 \qquad C_0=2\pi u_0\kappa^2 .
 \label{eq:density-hamiltonian}
\end{equation}
The centering by a traceless one-particle matrix therefore uses fixed
neutral particle number; it is not this same replacement in a general
grand-canonical sector.  For \(R^2=N_0/2\), \(C_0=gN_0/(4\pi)\).

The exact branch-exchange matrix \(\Gamma\) satisfies
\(U\Gamma=(\hat{\boldsymbol n}\cdot\boldsymbol\sigma)U\), which yields
\begin{equation}
 T_q=\Gamma A_q=A_q\Gamma,
 \qquad T_{00}=\frac{I_D}{\sqrt{4\pi}},\qquad
 \bar T_{00}=0,\qquad A_{00}=\frac{\Gamma}{\sqrt{4\pi}}.
 \label{eq:mass-charge-relation}
\end{equation}
The uniform density field drops out at half filling. The physical mass
monopole \(Q_{00}=M/\sqrt{4\pi}\), with \(M=f^\dagger\Gamma f\), remains
in all measurements.  Sampling density channels does not replace the mass
vertices in the observables.

\subsection{Auxiliary fields and the implemented time slice}
\label{sec:hs-timeslice}
For a centered density multipole \(\bar P_q\), a real Gaussian auxiliary field couples with an imaginary coefficient~\cite{Hubbard1959HS},
\begin{equation}
 e^{-\alpha^2\bar P_q^2/2}
 =\int\frac{dx}{\sqrt{2\pi}}\,e^{-x^2/2}e^{i\alpha x\bar P_q},
 \qquad \alpha=\sqrt{\Delta\tau u_0}.
 \label{eq:density-gaussian-hs}
\end{equation}
For a field history \(X=\{x_{qt}\}\), the one-particle propagator on slice \(t\) is
\begin{equation}
 B_t=e^{-\dt K/2}e^{iW_t}e^{-\dt K/2},\qquad
 W_t=\alpha\sum_{q\ne00}x_{qt}\bar T_q .
 \label{eq:hs-layer}
\end{equation}
To evaluate \(e^{iW_t}\), write \(W_t\) in positive/negative-energy blocks;
its diagonal blocks are equal and its two off-diagonal blocks are equal.
The fixed basis change that diagonalizes \(\Gamma\) gives two
\(N_0\times N_0\) blocks, \(W_\pm=W_{++}\pm W_{+-}\).
The signed permutation
\(J_{(k,m),(k',m')}=(-1)^{m+1/2}\delta_{kk'}\delta_{m',-m}\)
relates the two blocks through
\begin{equation}
 W_-=J W_+^*J^\dagger,\qquad
 e^{iW_-}=J\bigl(e^{iW_+}\bigr)^{\mathsf T}J^\dagger .
 \label{eq:density-half-block}
\end{equation}
Thus a single $N_0\times N_0$ Hermitian eigendecomposition suffices to construct $e^{iW_t}$.

One-body exponentials propagate Slater determinants through their single-particle matrices~\cite{Blankenbecler1981,Sugiyama1986PQMC}. For the normalized free-sea determinant with occupied columns \(Q_T\),
define \(L_\tau\Delta\tau=\Theta\).  The weight is
\begin{equation}
 \mathcal D[X]=\det[Q_T^\dagger B_{L_\tau}\cdots B_1Q_T],
 \qquad w[X]=e^{\Theta C_0}P[X]\mathcal D[X].
 \label{eq:weights-density}
\end{equation}
Here $P[X]$ is the product of the single-field priors.

The factor \(e^{\Theta C_0}\) cancels in normalized observables and acceptance ratios. The energy estimator includes the constant \(-C_0\) in the density representation. Because the density bilinears need not commute, their joint exponential can incur a local \(O(\dt^2)\) error, or \(O(\dt)\) at fixed projection length. Figure~\ref{fig:ed-dtau} tests this discretization against exact projection.

\subsection{Stabilized projection and a layer update}
\label{sec:sm-layer-update}
The total number of time slices is an even integer $L_\tau$, with
$L_\tau\dt=\Theta$. Thus $\Theta/2$ is projected on each side of a midpoint
measurement.
We use the Slater-determinant projector construction~\cite{Sugiyama1986PQMC,Sorella1989PQMC,AssaadEvertz2008} and the weights in Eq.~\eqref{eq:weights-density}.
The sampling target is $|w[X]|$, where $w[X]=P[X]\mathcal D[X]$ up to the
constant \(e^{\Theta C_0}\).
The determinant is single, not multiplied by a conjugate partner.

For an update of slice $t$, propagate the occupied columns from the two
boundaries to the adjacent cuts,
\begin{equation}
 R_t\sim B_{t-1}\cdots B_1Q_T,\qquad
 L_t^\dagger\sim Q_T^\dagger B_{L_\tau}\cdots B_{t+1}.
 \label{eq:sm-left-right}
\end{equation}
Here $\sim$ denotes stabilized bases for the same occupied subspaces, with
their determinant normalization factors tracked separately. The left columns
are obtained by propagation with adjoint matrices; no Hermiticity of the
density-HS slice is assumed. After each propagation the code uses thin QR,
chooses the triangular diagonal positive real, and accumulates its logarithmic
absolute determinant. This convention keeps every phase in the final overlap~\cite{White1989Stability,AssaadEvertz2008}.
Prefixes and suffixes are cached during alternating forward and reverse sweeps.

At the cut immediately after $B_t$ define
\begin{equation}
 Y_t=B_tR_t,\qquad O_t=L_t^\dagger Y_t,\qquad
 \rho_t=Y_tO_t^{-1}L_t^\dagger .
 \label{eq:sm-layer-rho}
\end{equation}
The code stores $G_t=I_D-\rho_t$ and evaluates the contraction using an
overlap solve. Replacing all fields on slice $t$ gives $B'_t$, $Y'_t$, and
$O'_t$. The unchanged normalization factors in Eq.~\eqref{eq:sm-left-right}
cancel, leaving
\begin{equation}
 r_{\mathcal D}=\frac{\det O'_t}{\det O_t},\qquad
 \rho'_t=Y'_t(O'_t)^{-1}L_t^\dagger .
 \label{eq:sm-overlap-ratio}
\end{equation}
LU factorization supplies the logarithmic determinant modulus and phase.
The right solve $Z O'_t=Y'_t$ supplies $\rho'_t=ZL_t^\dagger$ without a
slice inverse. After acceptance or rejection, the retained propagation is
orthonormalized once to update the sweep cache.

For the continuous centered-density implementation, the approximate determinant drift is
\begin{equation}
 F_q(X)=-\alpha\,\operatorname{Im}\operatorname{tr}
 [\bar T_q e^{+\dt K/2}\rho_t e^{-\dt K/2}],
 \qquad q\ne00 .
\end{equation}
The code contracts only the allowed entries of $\bar T_q$, with the kinetic factors absorbed into their coefficients. The Gaussian prior contributes $-x_{qt}$ to the drift. Writing $\bar x_q(X)=x_{qt}+b^2[F_q(X)-x_{qt}]/2$, the joint density is
\begin{equation}
 \log\mathcal Q(X'|X)=-\frac{N_q-1}{2}\log(2\pi b^2)
 -\frac{1}{2b^2}\sum_{q\ne00}[x'_{qt}-\bar x_q(X)]^2 .
 \label{eq:sm-proposal}
\end{equation}
The reverse density exchanges $X$ and $X'$ and uses the candidate mixed density. Equation~\eqref{eq:mh-main} in the main text uses both densities and the full Gaussian prior ratio. Thus the approximate force changes the proposal while the acceptance rule preserves detailed balance for $|w|$ at fixed $b$.

During initial thermalization, each chain updates its proposal width $b$ after the $i$th forward--reverse sweep pair according to
\begin{equation}
 b_{i+1}=b_i\exp[(a_i-0.574)/\sqrt{i}],
\end{equation}
We use $0.574$ as the target acceptance rate, motivated by the asymptotic result for Langevin proposals~\cite{Roberts1998MALA}. Here $a_i$ is the fraction of accepted layer proposals in that forward and reverse pair. Each chain tunes its own width, then freezes it for measurement.

The selection rules in Eq.~\eqref{eq:sm-mass-selection} bound assembly and drift contractions by \(O(D^2n_{\rm cut})\) per layer. A joint proposal requires one dense propagator and overlap calculation, giving \(O(D^3)\) total work per layer compared with \(O(N_qD^3)\) for dense reconstruction after each channel change.

\subsection{Midpoint contractions and physical observables}
\label{sec:sm-measurements}
At the midpoint $c=L_\tau/2$, let
$R_X\sim B_c\cdots B_1Q_T$ and
$L_X^\dagger\sim Q_T^\dagger B_{L_\tau}\cdots B_{c+1}$.
We use the index convention
\begin{equation}
 (\rho_X)_{IJ}=\langle f_J^\dagger f_I\rangle_X,\qquad
 \rho_X=R_X(L_X^\dagger R_X)^{-1}L_X^\dagger .
 \label{eq:sm-midpoint-rho}
\end{equation}
The contraction is generally complex and non-Hermitian. For arbitrary
one-particle matrices $A$ and $B$, Wick's theorem gives the ordered products~\cite{AssaadEvertz2008}
\begin{align}
 (f^\dagger Af)_X&=\operatorname{tr}(A\rho_X),\\
 [(f^\dagger Af)(f^\dagger Bf)]_X
 &=\operatorname{tr}(A\rho_X)\operatorname{tr}(B\rho_X)
 +\operatorname{tr}[A(I_D-\rho_X)B\rho_X].
 \label{eq:sm-wick}
\end{align}
For $A=B$, the first term is the algebraic square, not the absolute square.
All matrix traces and transpositions below use this same index convention.

The mass structure factor averages the second moment over a complete
angular multiplet,
\begin{align}
 S_\ell&=\frac{1}{2\ell+1}\sum_{\mu=-\ell}^{\ell}
 \langle Q_{\ell\mu}^2\rangle_\Theta\\
 &=\frac{1}{4\pi}\int d\Omega\,d\Omega'\,
 P_\ell(\hat{\bm n}\cdot\hat{\bm n}')
 \langle m_\Omega m_{\Omega'}\rangle_\Theta .
 \label{eq:mass-structure-factor}
\end{align}
Here $P_\ell$ is a Legendre polynomial. Each configuration contributes
\begin{equation}
 (S_\ell)_X=\frac{1}{2\ell+1}\sum_\mu
 \left\{[\operatorname{tr}(A_{\ell\mu}\rho_X)]^2
 +\operatorname{tr}[A_{\ell\mu}(I_D-\rho_X)A_{\ell\mu}\rho_X]\right\}.
 \label{eq:sm-Sl-estimator}
\end{equation}
In particular, $S_0=\langle M^2\rangle_\Theta/(4\pi)$.
The dimensionless correlation ratio is $\Rm=1-S_1/S_0$, formed after the
ensemble reweighting described below. These are total second moments.
The physical mass density is $m_\Omega/R^2$, so $S_0/N_0^2$ has the
size scaling of its spatially averaged squared mass at fixed density.

For shell $k$, define the diagonal vertex
\begin{equation}
 Z^{(k)}_{IJ}=\frac{\delta_{IJ}\delta_{k_I,k}}{2k}
 \begin{cases}-1,&s_I=+,\\+1,&s_I=-.\end{cases}
 \qquad
  (Z_k)_X=\operatorname{tr}[Z^{(k)}\rho_X] .
 \label{eq:sm-Zk}
\end{equation}
Phase reweighting this estimator gives $Z_k=(2k)^{-1}\sum_m
\langle n_{-km}-n_{+km}\rangle_\Theta$. It is the occupation contrast
between the two branches, equals one in the free sea, and samples momentum
$k/R$. The fermion scaling observable is the lowest-shell value $Z_1$.

\subsection{Complex reweighting and derived ratios}
\label{sec:sm-reweighting}
Let $\varphi_X=\arg\mathcal D[X]$. The phase and its mean are
\begin{equation}
 e^{i\varphi_X}=\frac{\mathcal D[X]}{|\mathcal D[X]|},\qquad
 \langle e^{i\varphi}\rangle_{|w|}
 =\frac{\sum_Xw[X]}{\sum_X|w[X]|} .
 \label{eq:sm-average-phase}
\end{equation}
A field-independent real normalization factor cancels in both ratios.
The finite-step expectation follows by full complex reweighting~\cite{Loh1990Sign,AssaadEvertz2008}:
\begin{equation}
 \langle O\rangle_{\Theta,\dt}=
 \frac{\langle e^{i\varphi_X}O_X\rangle_{|w|}}
 {\langle e^{i\varphi_X}\rangle_{|w|}} .
 \label{eq:sm-phase-reweight}
\end{equation}
At each parameter point, we sum the complex numerators and denominators separately over bins and independent chains before taking their ratio. We compute $\Rm=1-S_1/S_0$ from the resulting phase-reweighted structure factors and repeat these steps in each resample to retain their covariance.

\subsection{Time-step check against exact diagonalization}
\label{sec:sm-validation}
\begin{figure}[htbp]
\centering
\includegraphics[width=.62\linewidth]{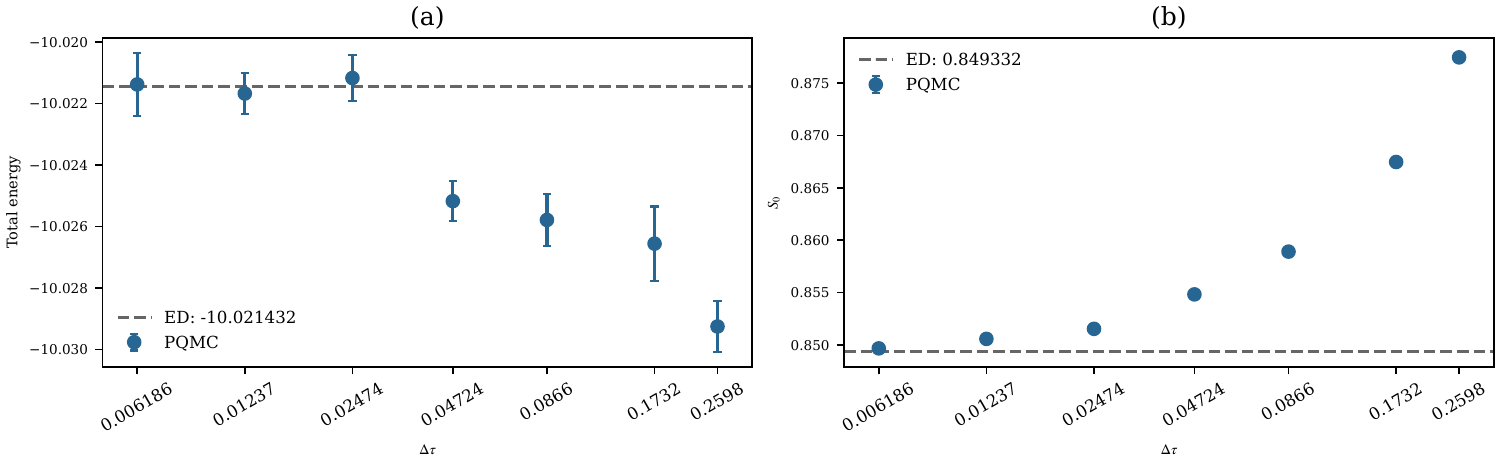}
\caption{Total energy (a) and scalar structure factor $S_0$ (b) versus time step at $N_0=6$ and $g=15.1$. Continuous-field density-HS PQMC uses an even number of time slices to keep the total projection length fixed. The horizontal axis gives the actual simulation step, and the dashed lines mark the exact projection values. Statistical errors are estimated by deleting one independent chain at a time.}
\label{fig:ed-dtau}
\end{figure}

At $N_0=6$ and $g=15.1$, we compare continuous-field density-HS PQMC with exact projection of the same Hamiltonian from the same free-sea state at $\tproj=0.6$. The seven displayed steps span $0.00619\le\Delta\tau\le0.260$. Coarser steps yield resolvable deviations from the exact projection values. As the step is reduced, both the total energy and the scalar structure factor $S_0$ approach their exact benchmarks [Fig.~\ref{fig:ed-dtau}].

\subsection{Phase diagnostics}
\label{sec:phase-comparison}
\begin{figure}[htb]
\centering
\includegraphics[width=.78\textwidth]{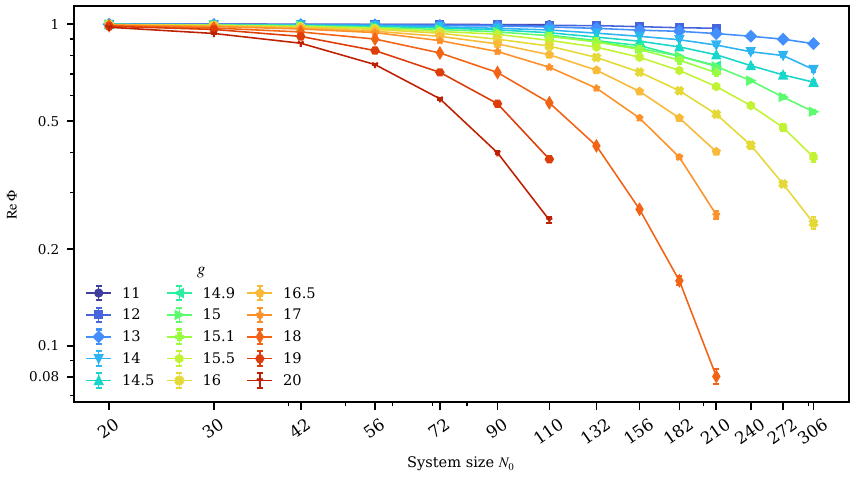}
\caption{Real part of the mean phase for the continuous-field density-HS simulations, with logarithmic scales on both axes. Each curve shows $\operatorname{Re}\Phi$ versus $N_0$ at fixed $g$, with $11\le g\le20$. Error bars are delete-one-chain jackknife standard errors.}
\label{fig:sm-fig1-phase}
\end{figure}

Write $\Phi=\langle e^{i\varphi}\rangle_{|w|}$ for the average in Eq.~\eqref{eq:sm-average-phase}.
At $\tproj=0.6$, $\operatorname{Re}\Phi$ is statistically resolved throughout the scan. Its smallest measured value is $0.0802(44)$, at $N_0=210$ and $g=18$ (Fig.~\ref{fig:sm-fig1-phase}).

\subsection{Crossings and slope ratios}
At fixed $\tproj$, the dimensionless mass ratio near the transition has the finite-size scaling form~\cite{ChenPhase2024,qinDuality2017}
\begin{equation}
 \Rm(g,R)=F(x)+R^{-\omega}G(x)+\cdots,\qquad x=(g-g_c)R^{1/\nu},
\end{equation}
where $\omega>0$ is the leading correction exponent. The crossing coupling $g_c^\star$ satisfies $\Rm(g_c^\star,R_1)=\Rm(g_c^\star,R_2)$. Expanding about $x=0$ and writing $\delta g=g_c^\star-g_c$ gives, to leading order,
\begin{equation}
 F'(0)\,\delta g\,(R_2^{1/\nu}-R_1^{1/\nu})
 =G(0)(R_1^{-\omega}-R_2^{-\omega}).
\end{equation}
For fixed $R_2/R_1>1$ and nonzero $F'(0)$, the crossing shift is therefore of order $R_1^{-1/\nu-\omega}$. At the crossing, both scaling arguments are of order $R_1^{-\omega}$. Differentiating the scaling form with respect to $g$ and evaluating at the crossing gives
\begin{equation}
 \frac{s_2}{s_1}=\left(\frac{R_2}{R_1}\right)^{1/\nu}
 \left[1+O(R_1^{-\omega})\right],\qquad
 s_i=\left.\partial_g\Rm(g,R_i)\right|_{g=g_c^\star}.
\end{equation}
The common amplitude $F'(0)$ cancels. For each size pair shown in Fig.~\ref{fig:local-crossings}(c) in the main text, we divide $\ln(s_2/s_1)$ by the logarithm of that pair's actual radius ratio $R_2/R_1$ to obtain $1/\nu_{\rm eff}$.

For each size we fit $\Rm(g,R)$ to a quadratic with equal weights over $14\le g\le16$. For each adjacent pair we find the crossing and evaluate both derivatives there using the same two quadratics. We repeat the fits, crossing search, and slope evaluation in each whole-chain bootstrap replicate. The error bars give the standard deviations over replicates with a unique crossing in the fit window.

\subsection{Uncertainties on the critical exponents}
The central estimates in Table~I of the main text come from the 76 available points with $110\le N_0\le306$, $12\le g\le18$, and $t_{\rm proj}=0.6$. We estimate the statistical error at this fixed window by drawing 400 bootstrap samples. At each $(N_0,g)$ we resample whole independent chains with replacement, drawing the original number of chains. Each draw carries the chain's complex phase sum and all observable numerator sums together. We sum these quantities over the sampled chains before forming the reweighted observables. In every replicate we repeat the inverse-variance mass-ratio fit for $g_c$ and $1/\nu$, then use its scaling coordinate in the fits of $S_0$ and $Z_1$. The point weights are held at their original jackknife estimates. The sample standard deviation across replicates gives $s_{\rm stat}$ and includes the propagation of the mass-ratio fit into both anomalous dimensions.

To quantify window sensitivity, we keep $N_{0,\min}=110$ and repeat the same fits for each combination of $N_{0,\max}=210,240,272,306$ and $g\in[12,18]$ or $[13,17]$, using all measured points in the selected window. For each exponent $p$, let $\hat p_0$ denote the unrounded primary estimate and $\hat p_w$ the estimate in window $w$. We compute
\[
 d_{\rm win}=\max_w |\hat p_w-\hat p_0|,\qquad
 h_p=\max(s_{\rm stat},d_{\rm win}).
\]
For all three exponents, the largest shift across fitting windows exceeds the fixed-window bootstrap error.
\begin{center}
\begin{tabular}{lrrrr}
\hline
Exponent & $\hat p_0$ & $s_{\rm stat}$ & $d_{\rm win}$ & Reported value \\
\hline
$1/\nu$    & 1.22609 & 0.01215 & 0.06282 & 1.23(7) \\
$\eta_\phi$ & 0.36819 & 0.00213 & 0.01156 & 0.37(2) \\
$\eta_\psi$ & 0.11915 & 0.00142 & 0.01173 & 0.119(12) \\
\hline
\end{tabular}
\end{center}
The largest shifts occur at $([13,17],272)$ for $1/\nu$, $([12,18],210)$ for $\eta_\phi$, and $([13,17],210)$ for $\eta_\psi$, where each pair specifies the coupling window and maximum size. Reported exponent uncertainties are rounded upward. These half-widths summarize the observed fit sensitivity at the simulated projection length and time steps; they carry no assigned confidence level for the asymptotic exponents. For $g_c$, the fixed-window bootstrap gives a statistical uncertainty of $0.015$. Across the eight fitting windows, the central estimate ranges from $15.058$ to $15.120$.

Table~\ref{tab:near-critical} tests the sensitivity to the two near-critical couplings included in the fits of Figs.~\ref{fig:results} and~\ref{fig:anomalous} in the main text. Including both groups shifts $1/\nu$ by $+0.00168$ and $\eta_\phi$ by $+0.00034$. The largest relative change occurs in the fermion channel: $\eta_\psi$ decreases by $0.00248(45)$, about one fifth of its quoted uncertainty, primarily upon including $g=15.1$. The uncertainty on this difference uses paired bootstrap replicates with identical draws at shared parameter points. All three shifts lie within the reported exponent uncertainties.

\begin{table}[htbp]
\caption{Sensitivity of the sphere fits to measurements at $g=14.9$ ($N_0=110,156,210$) and $15.1$ ($N_0=110,132,156,182,210$). Each row uses the same size and coupling limits, spline basis, and inverse-variance weights at shared points. For each selection we refit $g_c$ and $1/\nu$ from $r_{\rm mass}$ and propagate the resulting scaling coordinate into both anomalous-dimension fits. Parentheses give fixed-window statistical errors from 400 whole-chain bootstrap replicates. The last row is the main-text fit.}
\label{tab:near-critical}
\begin{ruledtabular}
\begin{tabular}{lrrrrr}
Near-critical groups included & Points & $g_c$ & $1/\nu$ & $\eta_\phi$ & $\eta_\psi$ \\
\hline
Neither & 68 & 15.076(17) & 1.224(12) & 0.3679(23) & 0.1216(15) \\
$14.9$ only & 71 & 15.077(17) & 1.224(12) & 0.3677(22) & 0.1213(14) \\
$15.1$ only & 73 & 15.068(15) & 1.226(12) & 0.3684(22) & 0.1194(14) \\
Both & 76 & 15.069(15) & 1.226(12) & 0.3682(21) & 0.1192(14) \\
\end{tabular}
\end{ruledtabular}
\end{table}

\subsection{Mass-ratio collapse with published estimates of $1/\nu$}
\begin{figure}[htbp]
\centering
\includegraphics[width=.95\linewidth]{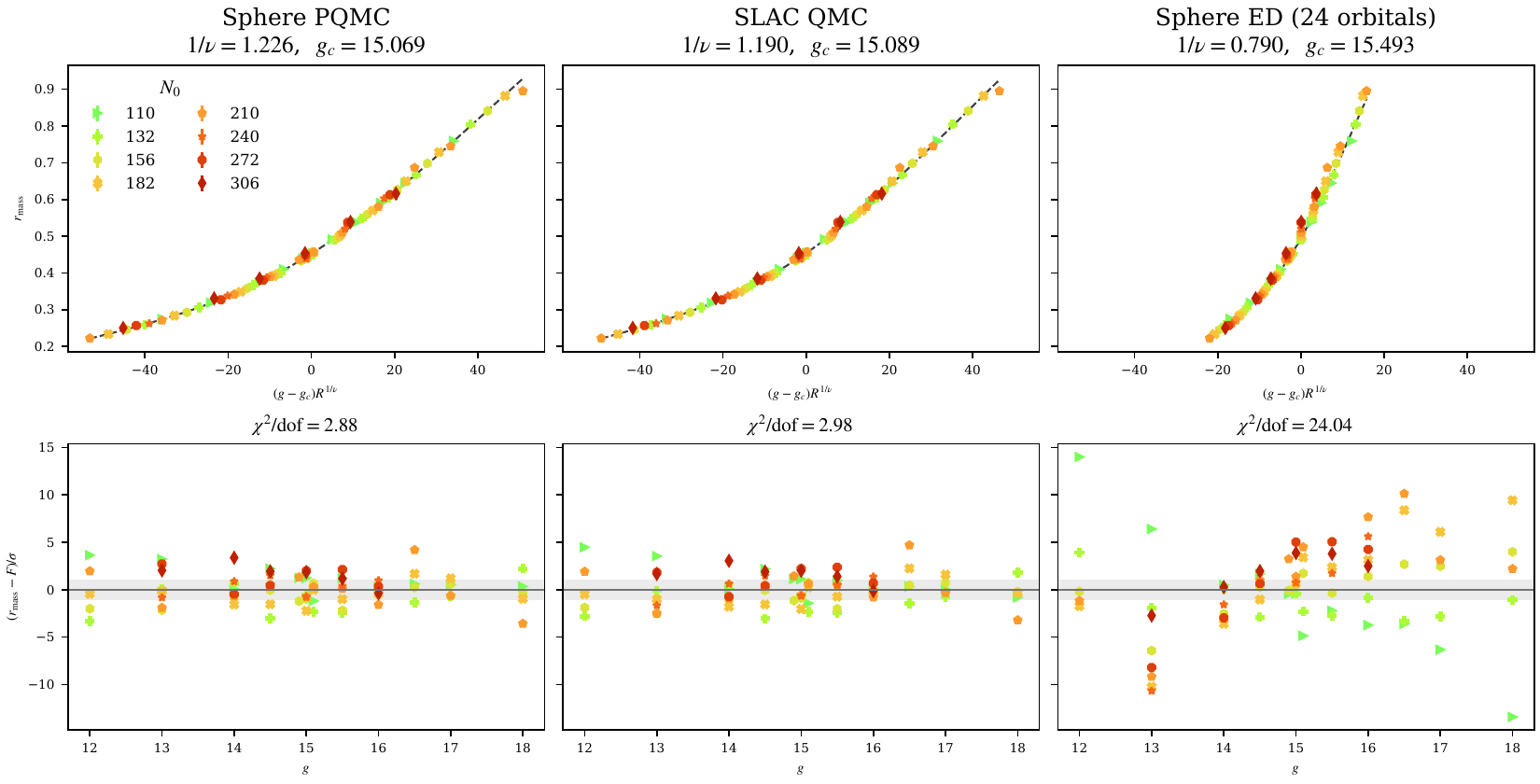}
\caption{Mass-ratio collapse for $N_0=110,132,156,182,210,240,272,306$ and $12\le g\le18$. From left to right, the value of $1/\nu$ comes from the spherical PQMC fit, SLAC QMC~\cite{Tabatabaei2022}, and spherical ED on 24 orbitals~\cite{Gao2026Sphere}. The upper panels show the data and fitted spline $F(x)$ with $x=(g-g_c)R^{1/\nu}$. The lower panels show signed residuals in units of the statistical standard error (shading, $\pm1$).}
\label{fig:sm-literature-collapse}
\end{figure}

Figure~\ref{fig:sm-literature-collapse} shows the collapse and signed residuals for three representative fits from Table~I of the main text. The freely fitted exponent gives $g_c=15.069$. Fixing $1/\nu$ to the SLAC values 1.19 and 1.18 shifts $g_c$ to 15.089 and 15.094, respectively. The ED values 0.79 and 0.65 shift it to 15.493 and 15.801.

\end{document}